\documentclass[onecolumn,showkeys,superscriptaddress,notitlepage]{revtex4-1}

\usepackage{graphicx}
\usepackage{amsmath,amsfonts,amssymb}
\usepackage{morefloats}
\usepackage{color}

\begin{document}

\title{How evolution affects network reciprocity in Prisoner's Dilemma}

\author{Giulio Cimini}
\email{g.cimini@math.uc3m.es}
\affiliation{Grupo Interdisciplinar de Sistemas Complejos (GISC), Departamento de Matem\'{a}ticas, Universidad Carlos III de Madrid, 28911 Legan\'{e}s, Madrid, Spain}
\author{Angel S\'{a}nchez}
\affiliation{Grupo Interdisciplinar de Sistemas Complejos (GISC), Departamento de Matem\'{a}ticas, Universidad Carlos III de Madrid, 28911 Legan\'{e}s, Madrid, Spain}
\affiliation{Instituto de Biocomputaci\'{o}n y F\'{i}sica de Sistemas Complejos (BIFI), Universidad de Zaragoza, 50018 Zaragoza, Spain}

\begin{abstract}
Cooperation lies at the foundations of human societies, yet why people cooperate remains a conundrum. 
The issue, known as network reciprocity, of whether population structure can foster cooperative behavior in social dilemmas has been addressed by many, but theoretical studies 
have yielded contradictory results so far---as the problem is very sensitive to how players adapt their strategy. 
However, recent experiments with the prisoner's dilemma game played on different networks have shown 
that humans do not consider neighbors' payoffs when making their decisions, and that the network structure does not influence the final outcome. 
In this work we carry out an extensive analysis of different evolutionary dynamics for players' strategies, 
showing that the absence of network reciprocity is a general feature of those dynamics that do not take neighbors' payoffs into account. 
Our results, together with experimental evidence, hint at how to properly model real people's behavior.
\end{abstract}
\maketitle

\section*{Introduction}

Cooperation and defection represent the two alternative choices behind social dilemmas \cite{Dawes1980}. 
Cooperative individuals contribute to the well-being of the community at their own expenses, whereas, defectors neglect doing so. 
Because of that cost of contribution, cooperators get lower individual fitness and thus selection favors defectors. This situation makes the emergence of cooperation a difficult matter. 
Evolutionary game theory \cite{MaynardSmith1973} represents a theoretical framework suitable to tackle the issue of cooperation among selfish and unrelated individuals. 
Within this framework, social dilemmas are formalized at the most basic level as two-person games, where each player can either choose to cooperate (C) or to defect (D). 
The {\it Prisoner's Dilemma} game (PD) \cite{Axelrod1984} embodies the archetypal situation in which mutual cooperation is the best outcome for both players, 
but the highest individual benefit is given by defecting. Mathematically, this is described by a matrix of payoffs (entries correspond to the row player's payoffs) of the form:
$$\begin{array}{c|cc}
	& \mbox{C}	& \mbox{D}	\\
\hline
\mbox{C}	& R	& S	\\
\mbox{D}	& T	& P	
\end{array}$$
so that mutual cooperation bears $R$ (reward), mutual defection $P$ (punishment), 
and with the mixed choice the cooperator gets $S$ (sucker's payoff) and the defector $T$ (temptation). 
The heart of the dilemma resides in the condition $T>R>P>S$: both players prefer the opponent to cooperate, but the temptation to cheat ($T>R$) 
and the fear of being cheated ($S<P$) pull towards choosing defection: according to Darwinian selection, cooperation extinction is then unavoidable 
\cite{Hofbauer1998}---a scenario known as the \emph{tragedy of the commons} \cite{Hardin1968}.

However, cooperation is widely observed in biological and social systems \cite{MaynardSmith1995}. 
The evolutionary origin of such behavior hence remains a key unsolved puzzle across several disciplines, ranging from biology to economics. 
Different mechanisms have been proposed as putative explanations of the emergence of cooperation \cite{Nowak2006}, 
including the existence of a social or spatial structure that determines the interactions among individuals---a feature known as \emph{network reciprocity}. 
In a pioneering work, Nowak and May \cite{Nowak1992} showed that the behavior observed in a repeated PD was dramatically different 
on a lattice than in a well-mixed population (or, in more physical terms, in a mean-field approach): 
in the first case, cooperators were able to prevail by forming clusters and preventing exploitation from defectors. 
Subsequently, many researchers, particularly from theoretical physics, devoted their attention to the problem of cooperation on complex networks, 
identifying many differences between structured and well-mixed populations \cite{Roca2009a} that by no means were always in favor of cooperation 
\cite{Hauert2004,SysiAho2005,Roca2009b}. The main conclusion of all these works is that this problem is very sensitive to the details of the system 
\cite{Szabo2007,Roca2009a,Roca2009b}, in particular to its evolutionary dynamics \cite{Hofbauer1998,Hofbauer2003} (\emph{i.e.}, the manner in which 
players adapt their strategy). On the experimental side, tests of the different models were lacking \cite{Helbing2010}, 
because the few available studies \cite{Cassar2007,Kirchkamp2007,Traulsen2010,Fischbacher2001,Suri2011} dealt only with very small networks.  
Network sizes such that clusters of cooperators could form have been considered only in recent large-scale experiments \cite{Grujic2010,Gracia2012b} 
with humans playing an iterated multiplayer PD, as in the theoretical models. The outcome of the experiments was that, when it comes to human behavior, 
the existence of an underlying network of contacts does not have any influence on the global outcome. 

The key observation to explain the discrepancy between theory and experiments is that most of the previous theoretical studies have been building on 
evolutionary dynamics based on payoff comparison \cite{Hofbauer2003,Roca2009a}. While these rules are appropriate to model biological evolution 
(with the payoff representing fitness and thus reproductive success), they may not apply to social or economic contexts---where individuals are aware 
of others' actions but often do not know how much they benefit from them. Also when the latter information is available, recent analysis \cite{Grujic2014} 
of experimental outcomes \cite{Fischbacher2001,Traulsen2010,Grujic2010,Gracia2012b} show that humans playing PD or Public Good games do not base 
their decisions on others' payoffs. Rather, they tend to reciprocate the cooperation that they observe, being more inclined to contribute  the more their partners do. 
The independence on the topology revealed in \cite{Gracia2012b} can be therefore seen as a consequence of this kind of behavior \cite{Gracia2012}. 
Notably, absence of network reciprocity has also been observed in theoretical studies based on Best Response dynamics (an update rule that, as we will see, 
is independent on neighbors' payoffs) \cite{Roca2009c} and in a learning-based explanation of observed behaviors \cite{Cimini2013}. 
This suggests that the absence of network reciprocity in the iterated PD may be general for any evolutionary dynamics that does not take neighbors' payoffs into account. 

In this paper we aim specifically at shedding light on this point. In order to do so, we develop and study an agent-based model of a population of individuals, 
placed on the nodes of a network, who play an iterated PD game with their neighbors (a setting equivalent to that of recent experiments 
\cite{Traulsen2010,Grujic2010,Gracia2012b}) and whose strategies are subject to an evolutionary process. The key point in this work is that we consider a large set 
of evolutionary dynamics, representing most of the alternatives that have been proposed so far to implement the strategy updating process. 
At the same time, we consider a large set of population structures, covering most of the studied models of complex networks. 
In this way we are able to make, on the same system, a quantitative comparison of the different evolutionary dynamics, 
and check the presence of network reciprocity in the different situations. 
At the end we show that the absence of network reciprocity is a general consequence of evolutionary dynamics which are not based on payoff comparison.

\section*{Model}

As we already mentioned, we consider a population of $N$ individuals, placed on the nodes of a network and playing an iterated PD game with their neighbors. 
During each round $t$ of the game, each player chooses to undertake a certain action (C or D) according to her strategy profile, then plays a PD game with her $k$ neighbors 
(the selected action remains the same with all of them) and finally receives the corresponding payoff $\pi^{(t)}$. 
Each player's strategy is modeled stochastically by the probability $p^{(t)}\in [0,1]$ of cooperating at round $t$. 
Strategies are subject to an evolutionary process, meaning that every $\tau$ rounds players may update their probabilities of cooperating according to a particular rule. 
Note that each player starts with an initial probability of cooperating $p^{(0)}$ drawn from a uniform distribution $Q[p^{(0)}]=\mathcal{U}[0,1]$. This results in an initial fraction 
of cooperators $c_0$ equal to $1/2$---a value close to what is observed in the experiments \cite{Grujic2010,Gracia2012b}, 
and otherwise representing our ignorance about the initial strategy of the players. In any event, our results remain valid for any (reasonable) form of the distribution $Q[p^{(0)}]$.

\begin{figure*}
\includegraphics[width=0.5\textwidth]{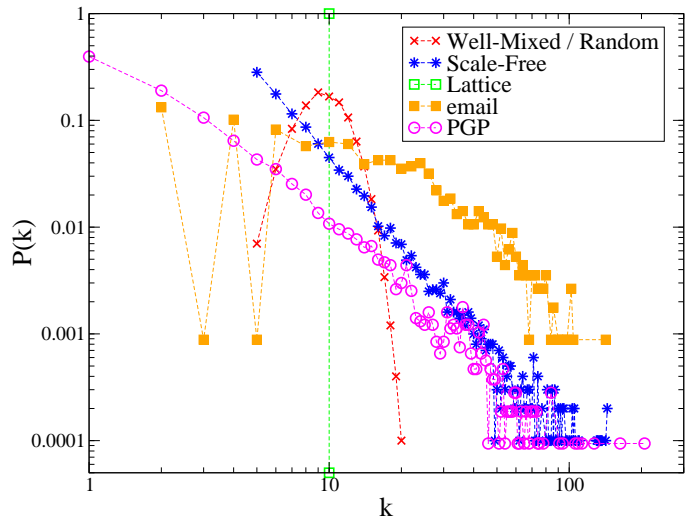}
\caption{Degree distributions $P(k)$ of the considered networks. Note that for the well-mixed case the network is dynamic but $P(k)$ does not change, 
and thus is identical to the $P(k)$ of random static networks.}\label{fig.degree}
\end{figure*}

We consider different parameterizations for the PD game, \emph{i.e.}, different intensities of the social dilemma. While we leave $R=1$ and $P=0$ fixed, 
we take $T$ values in the range $(1,2)$, and $S$ values in the range $(-1,0]$ (note that $S=P=0$ corresponds to the ``weak'' PD).
More importantly, we take into account different patterns of interactions among the players. 
These include the ``well-mixed'' population, represented by an Erd\"{o}s-R\'{e}nyi random graph with average degree $m$, rewired after each round of the game 
(which we indicate as \emph{well-mixed}), as well as static networks (all with average degree $m$): Erd\"{o}s-R\'{e}nyi random graphs (\emph{random}), scale free random networks 
with degree distribution $P(k)\sim k^{-3}$ (\emph{scale-free}) and regular lattice with periodic boundary conditions---where each node is linked to its $k\equiv m$ 
nearest neighbors (\emph{lattice}). Finally, we include two real instances of networks, the first given by the e-mail interchanges between members of the Univerisity Rovira i Virgili 
in Tarragona (\emph{email}) \cite{Guimera2003}, and the second being the giant component of the user network of the Pretty-Good-Privacy algorithm 
for secure information interchange (\emph{PGP}) \cite{Boguna2004}. The degree distributions of all these networks are reported in Figure \ref{fig.degree}. 
In simulations, we build the artificial networks using $N=1000$ and $m=10$ \cite{footB}, whereas, for the two real networks it is $N=1133$, $m=19.24$ (\emph{email}) and $N=10679$, $m=4.56$ (\emph{PGP}). 

The original and most important aspect of this study is that we consider a large variety of evolutionary dynamics for players to update their strategies. 
In addition, we remark that our study is more general than most of those encountered in the literature---where only pure strategies (\emph{i.e.}, playing always C or D) are considered. 
In our framework of mixed strategies, pure strategies can arise as special (limit) cases, when for a player $p$ becomes equal to 0 or 1.
\newline

We start by exploring a set of rules of imitative nature, representing a situation in which bounded rationality or lack of information force players to copy (imitate) 
others' strategies \cite{Schlag1998}. These rules are widely employed in the literature to model evolutionary dynamics. Here we consider the most notorious ones, 
in which a given player $i$ may adopt a new strategy by copying the probability of cooperating $p$ from another player $j$, which is one of the $|k_i|$ neighbors of $i$. 

\emph{Proportional Imitation} \cite{Helbing1992}---$j$ is chosen at random, and the probability that $i$ imitates $j$ depends on the difference between the payoffs 
that they obtained in the previous iteration of the game:
$$\mathcal P\left\{p_j^{(t)}\rightarrow p_i^{(t+1)}\right\}=\begin{cases}
	(\pi_j^{(t)}-\pi_i^{(t)})/\Phi_{ij}&\mbox{if $\pi_j^{(t)}>\pi_i^{(t)}$}\\
	0&\mbox{otherwise}
	\end{cases}$$
with $\Phi_{ij}=\max(k_i,k_j)[\max(R,T)-\min(P,S)]$ to have $\mathcal P\{\cdot\}\in[0,1]$. 
This updating rule is known to bring (for a large, well-mixed population) to the evolutionary equation of the replicator dynamics \cite{Schlag1998}.

\emph{Fermi rule} \cite{Szabo1998}---$j$ is again chosen at random, but the imitation probability depends now on the payoff difference according to the Fermi distribution function:
$$\mathcal P\left\{p_j^{(t)}\rightarrow p_i^{(t+1)}\right\}=\frac{1}{1+\exp[-\beta\,(\pi_j^{(t)}-\pi_i^{(t)})]}$$
Note that mistakes are possible under this rule: players can imitate others who are gaining less. 
The parameter $\beta$ regulates selection intensity, and is equivalent to the inverse of noise in the update rule. 

\emph{Death-Birth rule} (inspired by Moran dynamics \cite{Moran1962})---player $i$ imitates one of her neighbors $j$, or herself, with a probability proportional to the payoffs
$$\mathcal P\left\{p_j^{(t)}\rightarrow p_i^{(t+1)}\right\}=\frac{\pi_j^{(t)}-\psi}{\sum_{k\in \mathcal{N}_i^*}\pi_k^{(t)}-\psi}$$
where $\mathcal{N}_i^*$ is the set including $i$ and her neighbors and $\psi=\max_{j\in \mathcal{N}_i^*}(k_j)\min(0,S)$ to have $\mathcal P\{\cdot\}\in[0,1]$. 
As with the Fermi rule, mistakes are allowed here. 

\emph{Unconditional Imitation} or ``Imitate the Best'' \cite{Nowak1992,footUI}---under this rule each player $i$ imitates the neighbor $j$ with the largest payoff, 
provided this payoff is larger than the player's:
$$\mathcal P\left\{p_j^{(t)}\rightarrow p_i^{(t+1)}\right\}=1\quad\mbox{ if }j:\pi_j^{(t)}=\max_{k\in \mathcal{N}_i^*}\pi_k^{(t)}$$
Note that while the first three rules are stochastic, Unconditional Imitation leads to a deterministic dynamics. 

\emph{Voter model} \cite{Holley1975}---$i$ simply imitates a randomly selected neighbor $j$. 
Differently from the other imitative dynamics presented here (in which the imitation mechanism is based on the payoffs obtained in the previous round of the game), 
the Voter model is not based on payoff comparison, but rather on social pressure: players simply follow the social context without any strategic consideration \cite{Fehr2000,Vilone2012}. 
\newline

We also consider two evolutionary dynamics which go beyond pure imitation and are innovative, allowing extinct strategies to be reintroduced in the population 
(whereas imitative dynamics cannot do that). As we will see, neither of these rules (nor the Voter model) make use of the information on others' payoffs. 

\emph{Best Response} \cite{Matsui1992,Blume1993}---This rule has been widely employed in economic contexts, embodying a situation in which players are rational enough 
to compute the optimum strategy, \emph{i.e.}, the ``best response'' to what others did in the last round. More formally, at the end of each round $t$ a given player $i$ uses $x_i^{(t)}$ 
(the fraction of neighbors who cooperated in $t$) to compute the payoffs that she would have obtained by having chosen action C or D, respectively:
$$E\left[\pi_i^{(t)}(C)\right]=R\,x_i^{(t)}+S\,(1-x_i^{(t)})\qquad;\qquad E\left[\pi_i^{(t)}(D)\right]=T\,x_i^{(t)}+P\,(1-x_i^{(t)})$$
The quantity to increase is then:
$$E\left[\pi_i^{(t)}\right]=p_i^{(t)}\,E\left[\pi_i^{(t)}(C)\right]+(1-p_i^{(t)})\,E\left[\pi_i^{(t)}(D)\right]$$
The new strategy $p_i^{(t+1)}$ is picked among $\{p_i^{(t)},\,p_i^{(t)}-\delta,\,p_i^{(t)}+\delta\}$ (where $\delta$ is the ``shift'') as the one that brings to the highest $E[\pi_i^{(t)}]$ 
(and satisfies $0\le p_i^{(t+1)}\le 1$). Note that we do not use exhaustive best response here (which consists in choosing $p_i^{(t+1)}$ as the value of $p$ that maximize $E[\pi_i^{(t)}]$) as it leads immediately 
to the Nash equilibrium of the PD game ($p_i=0$ $\forall i$). Best Response belongs to a family of strategy updating rules known as \emph{Belief Learning} models, 
in which players update beliefs about what others will do according on accumulated past actions, and then use those beliefs to determine the optimum strategy. 
Best Response is a limit case that uses only last round’s actions to determine such optimum. We chose to restrict our attention to Best Response for three main reasons. 
1) it allows for a fair comparison with the other updating rules, that only rely on last round’s information; 2) in the non-exhaustive formulation of Best Response, history is held 
in the actual values of the parameter $p$; 3) in PD the Nash equilibrium is full defection, hence at the end the system collapses to this state for any information used to build beliefs about others' actions. 

\emph{Reinforcement Learning}---\cite{Bush1955,Macy2002,Izquierdo2008,Cimini2013}. Under this rule, a player uses her experience to choose or avoid certain actions 
based on their consequences: choices that met aspirations in the past tend to be repeated in the future, whereas, choices that led to unsatisfactory outcomes tend to be avoided. 
This rule works as follows. First, after each round $t$, player $i$ calculates her \emph{stimulus} $s_i^{(t)}$ as 
$$s_i^{(t)}=\frac{\pi_i^{(t)}/k_i-A_i^{(t)}}{\max\{|T-A_i^{(t)}|,|R-A_i^{(t)}|,|P-A_i^{(t)}|,|S-A_i^{(t)}|\}}$$
where $A_i^{(t)}$ is the current \emph{aspiration level} of player $i$, and normalization assures $|s_i^{(t)}|\le1$. 
Second, each player updates her strategy as:
	$$p_i^{(t+1)}=
	\begin{cases}
	p_i^{(t)}+\lambda s_i^{(t)}\,(1-p_i^{(t)})&\mbox{if $s_i^{(t)}>0$}\\
	p_i^{(t)}+\lambda s_i^{(t)}\,p_i^{(t)}&\mbox{if $s_i^{(t)}<0$}
	\end{cases}$$
where $\lambda\in(0,1]$ is the learning rate---low and high $\lambda$s representing slow and fast learning, respectively (hence for simplicity we use $\tau=1$ in this case). 
Finally, player $i$ can adapt her aspiration level as $A_i^{(t+1)}=(1-h)\,A_i^{(t)}+h\,\pi_i^{(t)}/k_i$, where $h\in[0,1)$ is the adaptation (or habituation) rate. 
Note that, when learning, players rely only on the information about their own past actions and payoffs.

\section*{Results and Discussion}

Here we report the results of the extensive simulation program for the model described above. In the following discussion, we focus our attention 
on the particular evolutionary dynamics employed, as well as on the specific network topology describing the interaction patterns among the players. 
We study the evolution of the level of cooperation $c$ (\emph{i.e.}, the fraction of cooperative players in each round of the game), as well as 
the stationary probability distribution of the individual strategies (\emph{i.e.}, the parameters $p$) among the population. We will show results relative to the case $\tau=10$ 
({\emph{i.e.}, we update players' strategy every ten rounds), yet we have observed that the particular value of $\tau$ influences the convergence time 
of the system to its stationary state, but does not alter its qualitatively characteristics. Also, we will report examples for two sets of game parameters 
($T=3/2,\,S=-1/2$ and $T=3/2,\,S=0$) but our findings are valid for the whole range studied---the main differences appearing between 
the ``strong'' and ``weak'' version of the PD game. Finally, we will show results averaged over a low number of realizations because, as the experiments show, 
the absence of network reciprocity is observed for single realizations, and hence it should be recovered from the model in the same manner.

\subsection*{Imitative dynamics} 

Results for the different imitation-based strategy updating rules are reported in Figures\ \ref{fig.PI_FR_DB} and \ref{fig.UI_VM_BR}. As plots clearly show, 
the final level of cooperation in this case depends heavily on the population structure, and often the final state is full defection (especially for the well-mixed case). 
The fact that these updating are imitative and not innovative generally leads, for an individual realization of the system, to a very low number 
of strategies $p$ (often just one) surviving at the end of the evolution. However, the surviving strategies are indeed different among independent realizations 
(but when $p\rightarrow0$, \emph{i.e.}, the final outcome is a fully defective state). This points out to the absence of strategies that are evolutionarily stable. 

The easiest situation to understand is perhaps given by employing the Voter model as the update rule: since there is no mechanism here to increase the payoffs, 
the surviving strategy is just randomly selected among those initially born in the population, and thus the average cooperation level is determined by $Q[p^{(0)}]$ 
(the probability distribution of the initial parameters $p$). This happens irrespectively of the particular values of $T$ and $S$ and, more importantly, 
of the specific topology of the underlying social network. A similar situation is observed with the Fermi rule for low $\beta$ (high noise). Indeed, in this case 
errors are frequent, so that players copy the strategies of others at random and $c$ remains close on average to its initial value $c_0$. 
The opposite limit of high $\beta$ (small noise, the case reported in the plots) corresponds instead to errors occurring rarely, meaning that players always copy the strategy 
of others who have higher payoffs. In the majority of cases, for the strong PD this leads to a fully defective final outcome. The exception is given by games played on 
network topologies with broad degree distribution, where cooperation may thrive at a local scale (resulting in a small, non-zero value of $c$) because of the presence of 
hubs---see below for a more detailed discussion of this phenomenon. On the other hand, the weak PD showcases more diverse final outcomes: the stationary value 
of $c$ is higher than $c_0$ for scale-free topologies, and a non-zero level of cooperation arises also in static random graphs and lattices. 
Note that in general we observe that the stationary (non-zero) cooperation levels decrease/increase for increasing/decreasing values of the temptation $T$, 
however such variations do not alter qualitatively the picture we present here---for this reason, we only present results for $T=3/2$. 
Moving further, Proportional Imitation leads to final outcomes very similar to those of the Fermi Rule for high $\beta$, which makes sense as the two rules are very similar---the only difference 
being the form of the updating probability (linear in the payoff difference for Proportional Imitation, highly non-linear for the Fermi rule). 
The Death-Birth rule and Unconditional Imitation also bring to similar results, and this is also due to their similarity in preferentially selecting the neighbor with the highest payoff. 
For these latter two rules, cooperation emerges for games played on all kinds of static networks (\emph{i.e.}, it does not only for a well-mixed population), with a stationary value of $c$ 
which varies depending on the specific network topology and on the particular entries of the payoff matrix ($T$ and $S$). 

While explaining in detail the effects of a particular updating rule and of a given population structure is out of the scope of the present work, 
we can still gain qualitative insights on the system's behavior from simple observations. Here we discuss the case of networks with highly heterogeneous degree distribution, 
such as scale-free ones. These topologies are characterized by the presence of players with high degree (``hubs'') that generally get higher payoff than an average player's 
as they play more instances of the game (the average payoff being greater than 0). For a dynamics of imitative nature, hubs' strategy remains stable: they hardly copy 
their less-earning neighbors, who in turn tend to imitate the hubs. As a result, hubs' strategy spread locally over the network, and, if such strategy profile results in frequent cooperation, 
a stable subset of player inclined to cooperate can appear around these hubs---see, \emph{e.g.}, \cite{Gomez2007}. The same situation cannot occur in random or regular graphs, 
where the degree distribution is more homogeneous and there are no hubs with systematic payoff advantage. This phenomenon becomes evident with Proportional Imitation as the updating rule. 
Note that the subset of players around hubs loses stability if they can make mistakes (as with the Fermi rule); on the other hand, such stability is enhanced 
when the updating rule preferentially selects players with high payoffs (as with the Death-Birth rule and Unconditional Imitation), because hubs' strategy spreads more easily. 
The fact that in the latter two cases cooperation thrives also in lattices is instead related to the emergence of clusters of mutually connected players who tendentially cooperate, 
get higher payoff than the defectors at the boundary of the cluster exploiting them and can thus survive. 

We finally remark that, beyond all the particular features and outcomes of each imitative dynamics, the main conclusion of this analysis is that 
imitation based on payoff comparison does not lead to the absence of network reciprocity, and that the only updating rule whose behavior is not affected 
by the population structure is the Voter model (which however does not depend on payoffs).

\subsection*{Non-imitative dynamics} 

The first general remark about these evolutionary rules is that they allow extinct strategies to be reintroduced in the population; because of this, many strategy profiles survive 
at the end of each realization of the system (even when the dispersion of the parameters $p$ is small). Results for Best Response dynamics are reported in the left column 
of Figure\ \ref{fig.UI_VM_BR}. We recall that this way of updating the strategies is the most ``rational'' among those we are considering in this work, and is not based 
on comparing own payoffs with those of others. As a result, we see that for the strong PD the system converges towards full defection for any value of the temptation 
and for any population structure. Indeed, this outcome is the Nash equilibrium of PD games, that would have been obtained also by global maximization of the individual expected payoffs. 
Hence the specific value of $\delta$ (\emph{i.e.}, the amount by which the parameters $p$ can be shifted at each update) only influences the time of convergence to full defection, 
with higher $\delta$ causing simulations to get faster to $p_i=0$ $\forall i$. For this reason, we only show results for a particular value of $\delta$. 
For the weak PD we observe instead a semi-stationary, non-vanishing yet slightly decreasing level of cooperation---which is the consequence of actions C and D bringing to the same payoff 
when facing  a defector. Such cooperation level seems to depend on the network size (bigger networks achieve higher $c$), rather than on the network topology. 
In this sense, we can claim that evolution by Best Response features absence of network reciprocity. This conclusion is supported by the fact that the optimal choice for a PD game 
does not depend on what the others do or gain, and as a consequence the social network in which players are embedded must play no role.

The other non-imitative rule that we consider in this study is Reinforcement Learning. Results for this choice of the dynamics are shown in Figure\ \ref{fig.RL}. 
We start with the simplest assumption of aspiration levels $A$ remaining constant over iterations of the game. 
Here the most remarkable finding is that, in contrast to all other update schemes discussed so far, with this dynamics evolutionary stable mixed strategies do appear: 
the parameters $p$ tend to concentrate around some stationary, non-trivial values, that do not depend on the initial condition of the population, neither on the topology of the underlying network. 
Concerning cooperation levels, when aspirations $A$ are midway between the punishment and reward payoffs ($P<A<R$) we observe a stationary, non vanishing $c\in[0.3,0.4]$ 
(which is not far from what is observed in experiments). The specific value of $c$ does depend on the payoff's matrix entries $T$ and $S$, but not on the population structure. 
Note that the described behavior is robust with respect to the learning rate $\lambda$ \cite{footL}. We can thus assert that Reinforcement Learning represents another evolutionary dynamics 
which can explain the absence of network reciprocity. This happens because, for this choice of updating rule, players do not take into consideration others' actions nor payoffs 
when adjusting their strategy, and thus the patterns of social interactions become irrelevant. 
Additional evidence for the robustness of this updating scheme derives from the behavior observed for other aspiration levels, including dynamic ones. 
As a general remark, the final level of cooperation reached is higher for higher $A$. For instance, when $R<A<T$ an outcome of mutual cooperation does not meet players' aspirations, 
however an outcome of mutual defection is far less satisfactory and brings to a substantial increase of $p$ for the next round. 
Because of this feedback mechanisms, players' strategies tend to concentrate around $p=1/2$ (\emph{i.e.}, playing C or D with equal probability), which thus results in $c\simeq 1/2$, 
again irrespectively of the population structure. Leaving aside the questionable case of aspiration levels below punishment ($S<A<P$), we finally consider adaptive aspiration levels. 
What we observe in this case is that players' aspirations become stationary---with final values falling in the range $P<\bar{A}<R$---and that no topological effects are present 
(as with Best Response, bigger networks achieve slightly higher cooperation). 

Summing up, we observe absence of network reciprocity for the innovative dynamics considered here, which we recall are not based on payoff comparison. 
This again supports our assumption that such outcome derives from not taking into account the payoffs of the rest of the players.

\section*{Conclusion}

Understanding cooperation represents one of the biggest challenges of modern science. Indeed, the spreading of cooperation is involved in all major transitions of evolution 
\cite{MaynardSmith1995}, and the fundamental problems of the modern world (resource depletion, pollution, overpopulation, and climate change) are all characterized by the tensions 
typical of social dilemmas. 
This work has been inspired by the experimental findings \cite{Grujic2010,Gracia2012b} that network reciprocity is not a mechanism to promote cooperation within humans playing PD. 
We aimed at identifying the evolutionary frameworks that are unaffected by the interaction patterns in the population, that are thus able to properly model real people behavior. 
To this end, we have considered several mechanism for players to update their strategy---both of imitative nature and innovative mechanisms, 
as well as rules based on payoff comparison and others based on non-economic or social factors. 
We stress that this is a very relevant point, as for the first time to our knowledge we are providing an extensive comparison of payoff-based and non-payoff based evolutionary dynamics 
on a wide class of networks (representing population structures). Our research suggests that absence of network reciprocity is a general feature of evolutionary dynamics 
in which players do not base their decisions on others' well-being. Note that the evolutionary dynamics we excluded as possible responsibles for how people behave 
are difficult to justify for humans and in a social context, because they assume very limited rationality that only allows to imitate others. 
Indeed, analysis of experimental outcomes \cite{Grujic2014} point out that humans playing PD do not base their decisions on others' payoffs. 
We thus believe that the present work provides a firm theoretical support for these experimental results, and allows to conclude that many of the evolutionary dynamics, 
based on payoff comparison, used in theory and in simulations so far simply do not apply to the behavior of human subjects and, therefore, their use should be avoided.

\begin{acknowledgments}
This work was supported by the Swiss Natural Science Fundation through grant PBFRP2\_145872 and by Ministerio de Econom\'\i a y Competitividad (Spain) through grant PRODIEVO.
\end{acknowledgments}

\begin{figure*}
\includegraphics[width=0.3\textwidth]{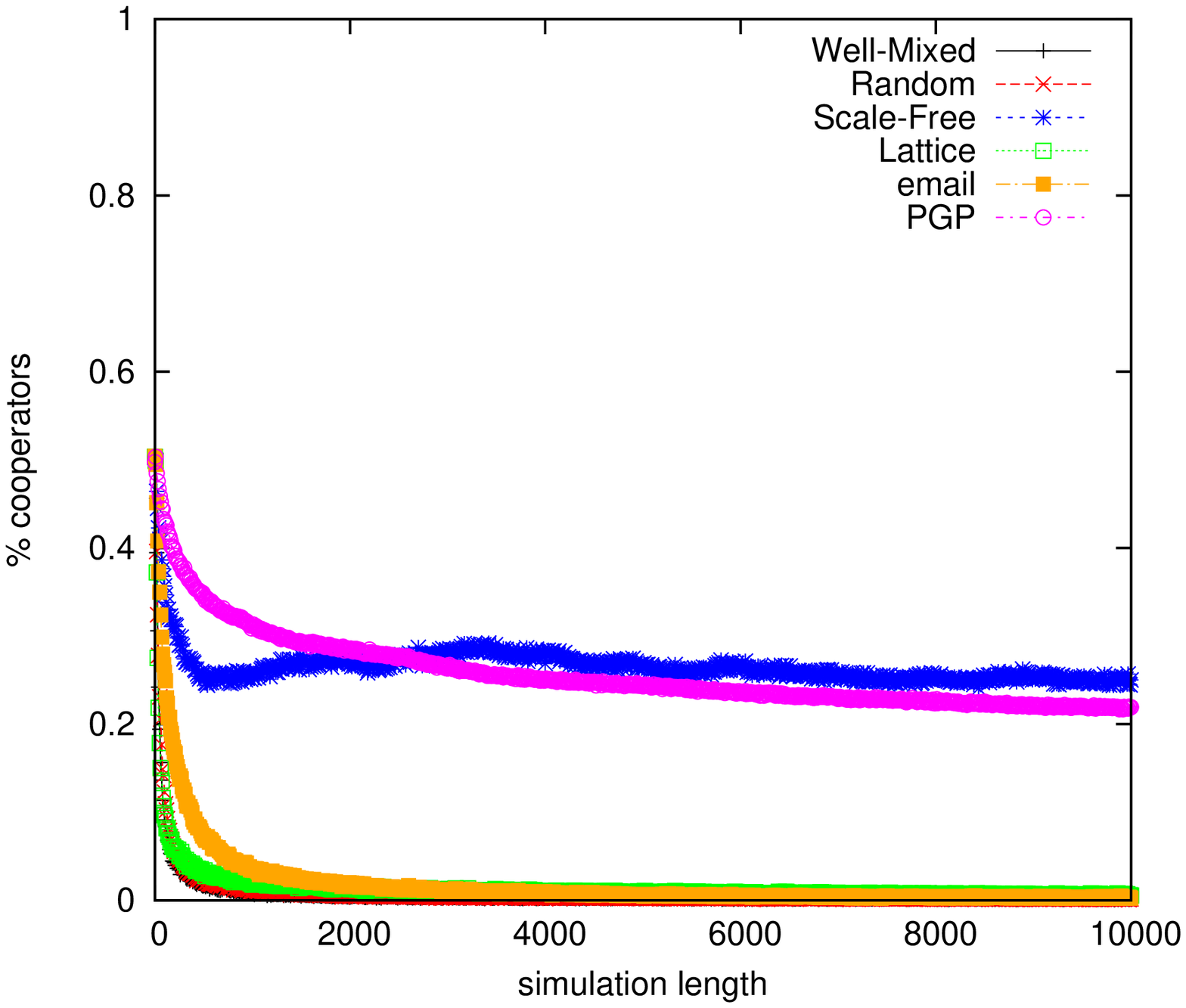}
\hspace{0.5cm}
\includegraphics[width=0.3\textwidth]{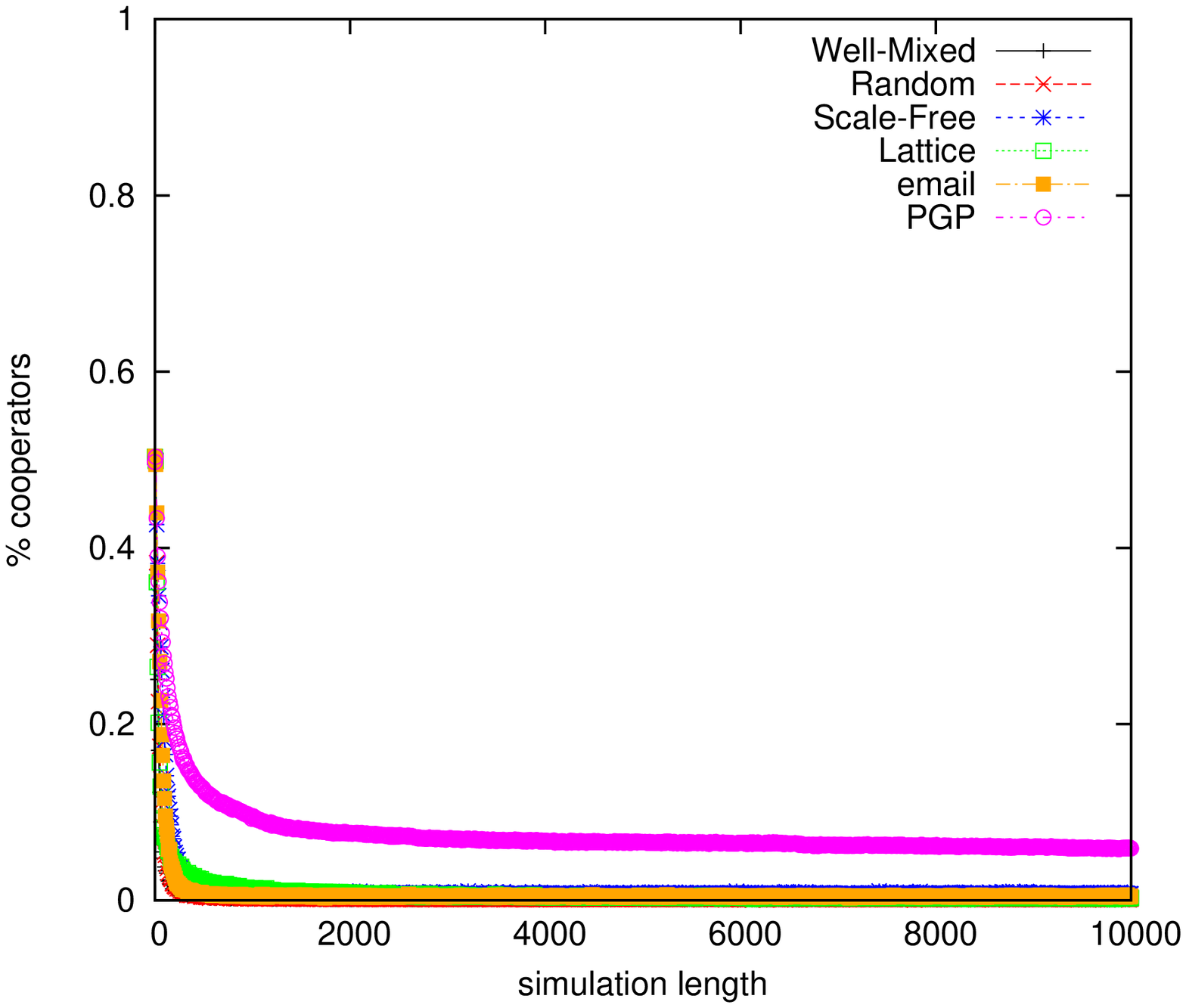}
\hspace{0.5cm}
\includegraphics[width=0.3\textwidth]{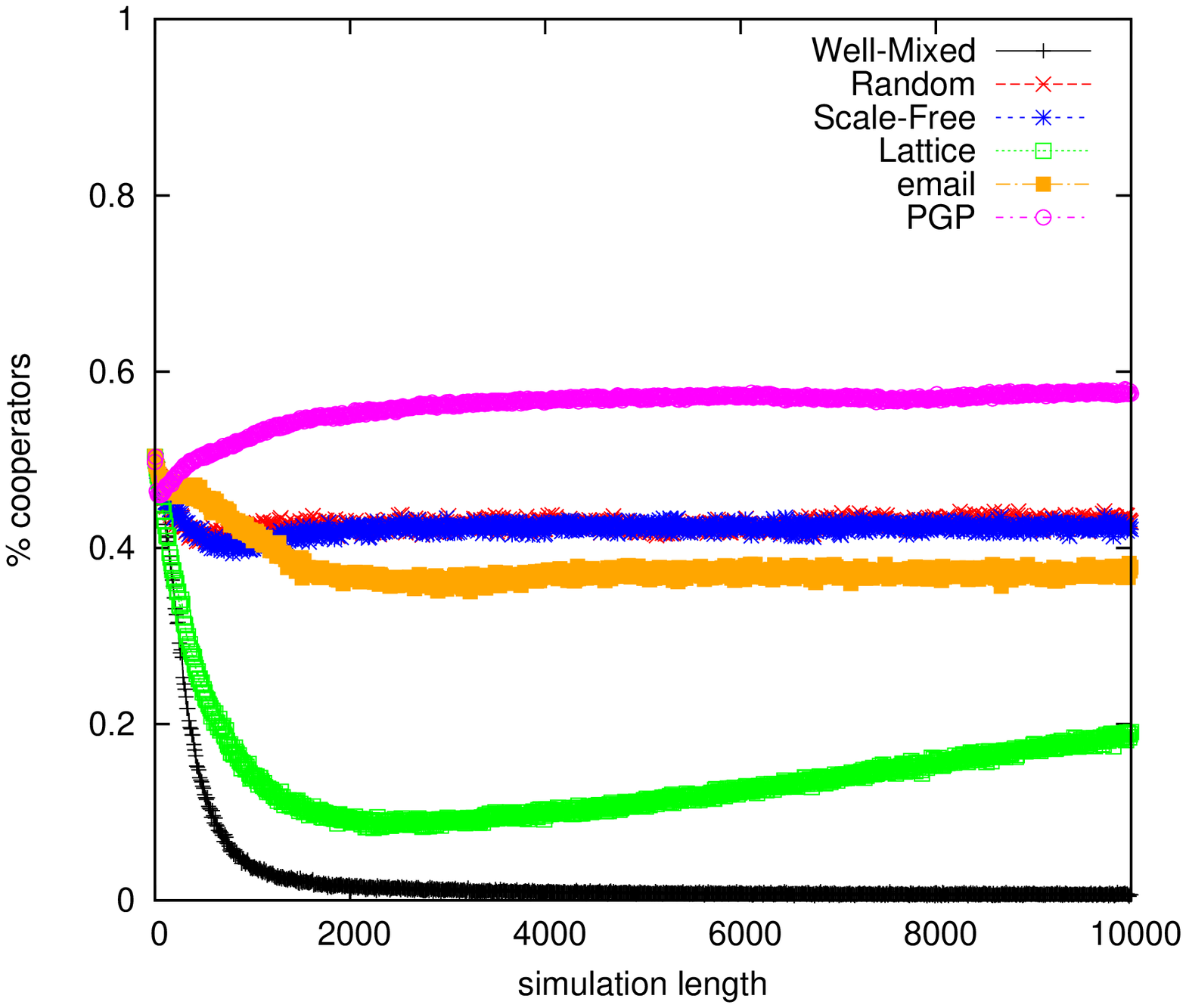}
\newline
\includegraphics[width=0.3\textwidth]{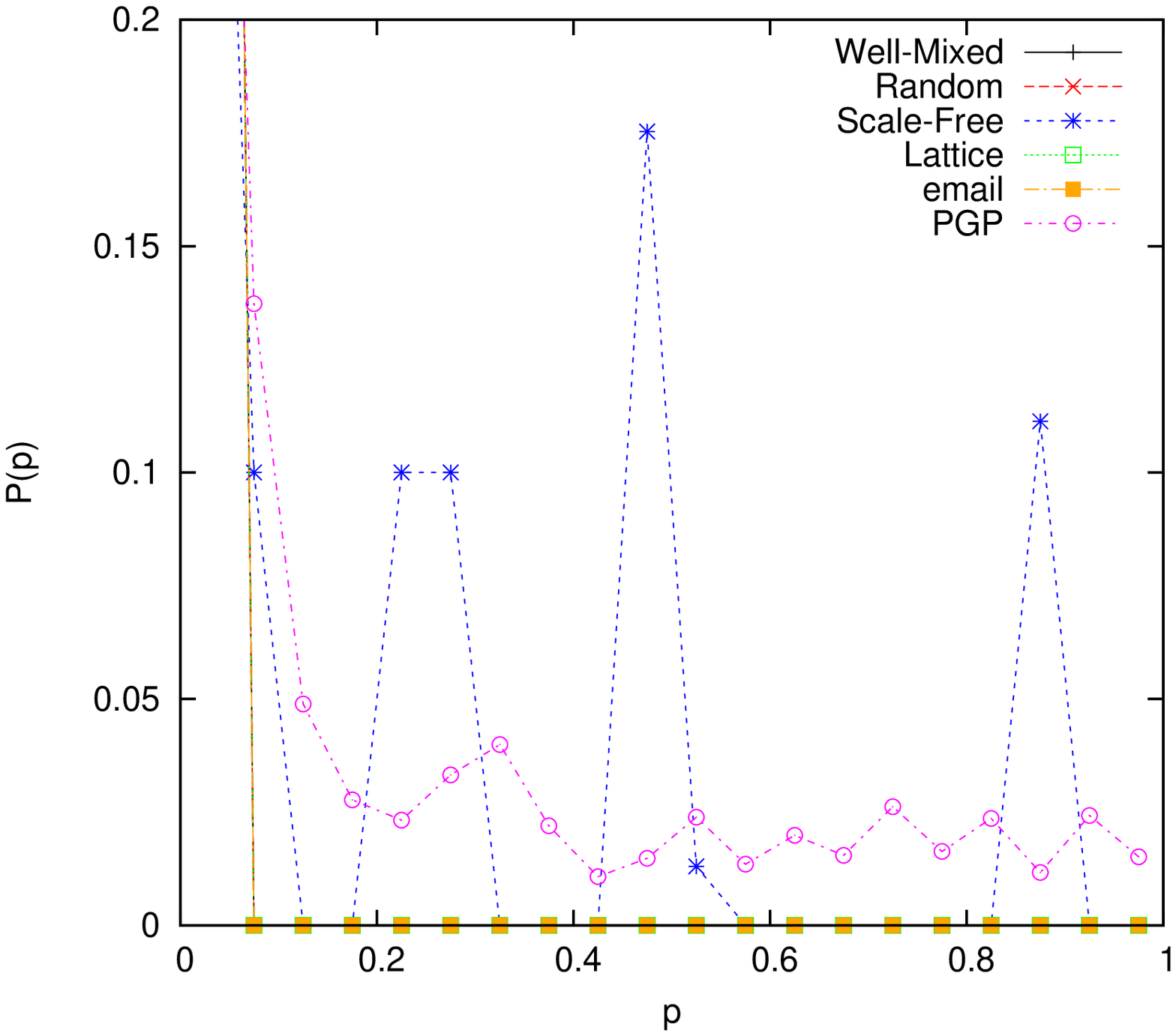}
\hspace{0.5cm}
\includegraphics[width=0.3\textwidth]{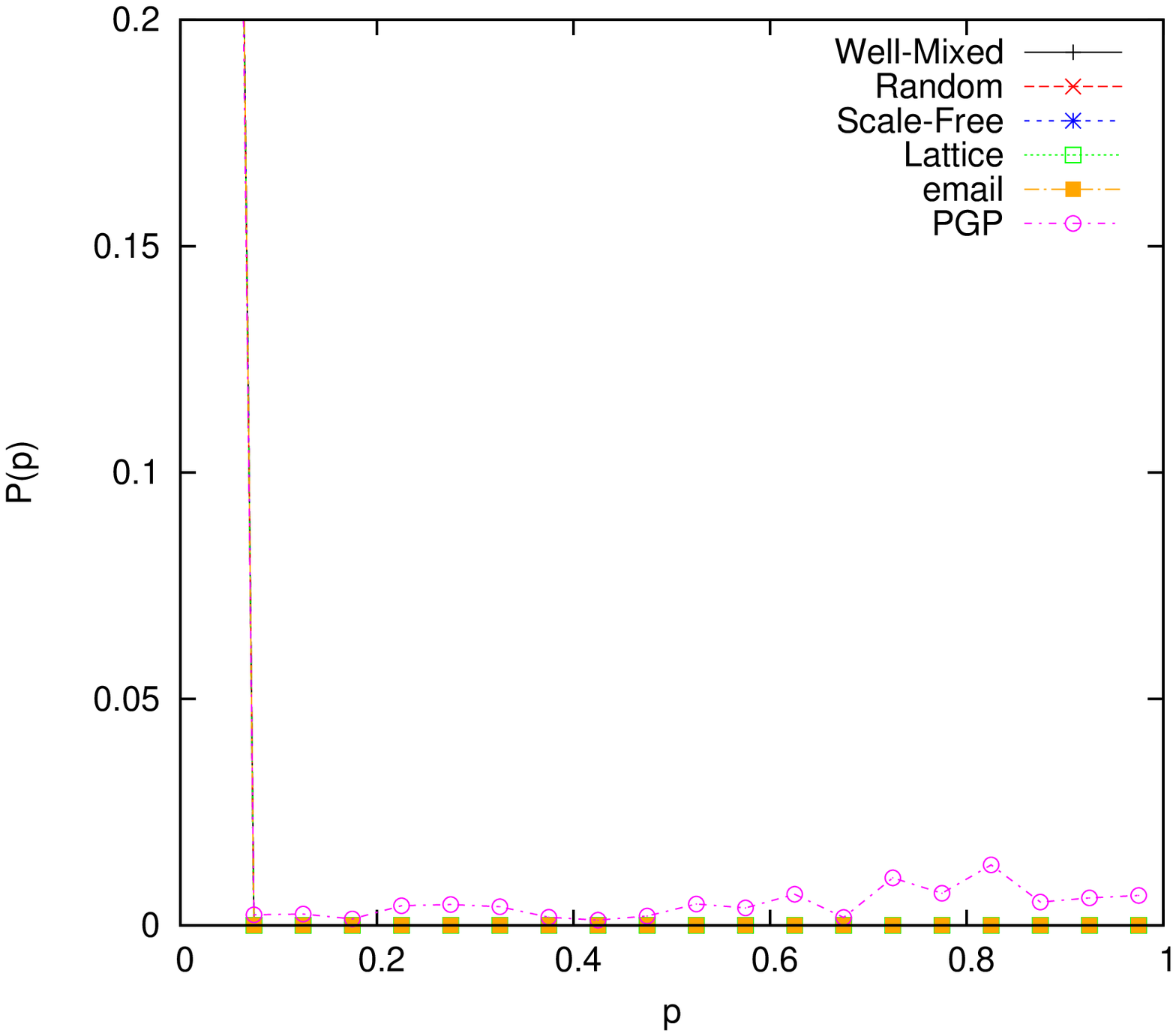}
\hspace{0.5cm}
\includegraphics[width=0.3\textwidth]{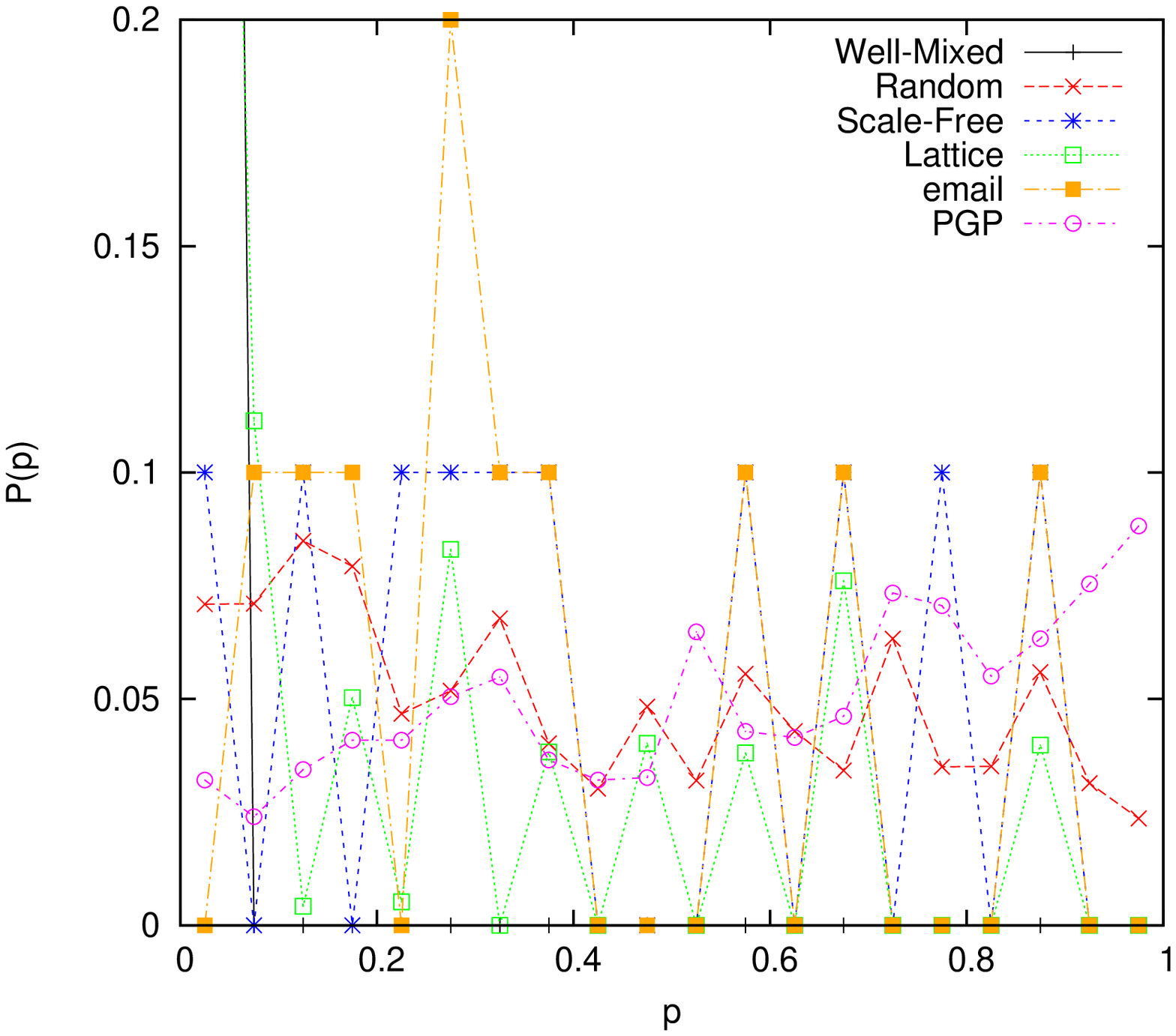}
\newline
\vspace{0.5cm}
\newline
\includegraphics[width=0.3\textwidth]{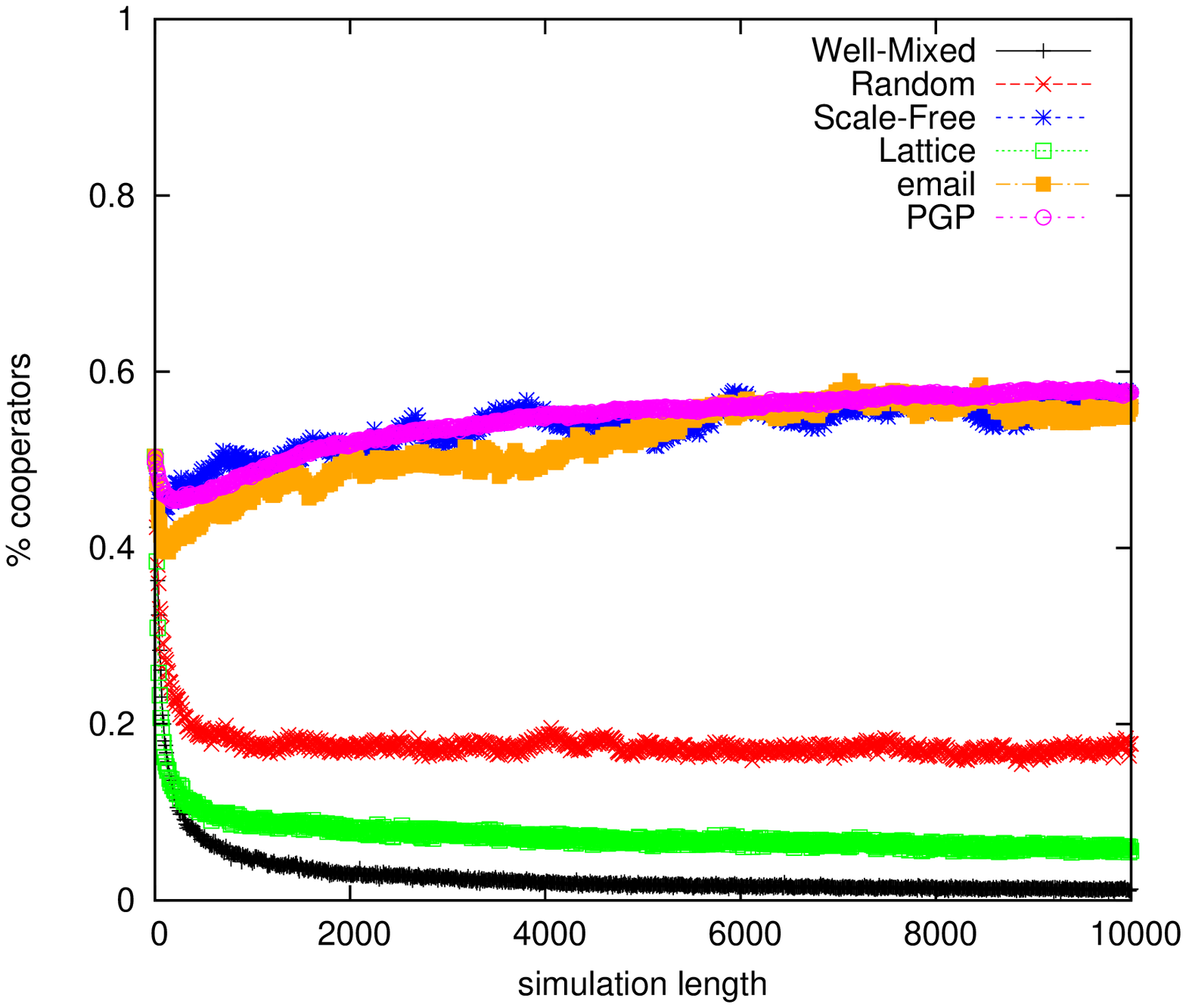}
\hspace{0.5cm}
\includegraphics[width=0.3\textwidth]{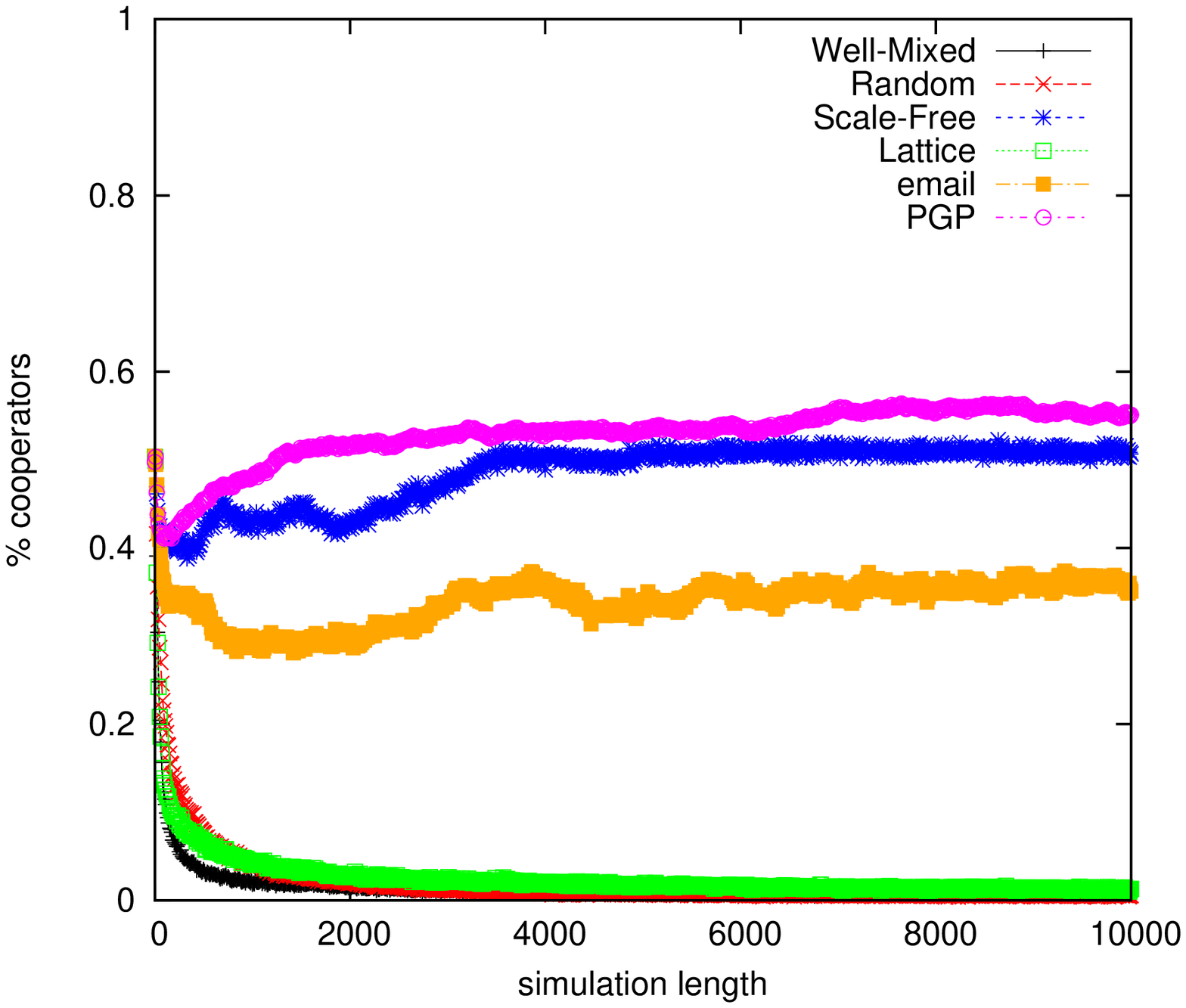}
\hspace{0.5cm}
\includegraphics[width=0.3\textwidth]{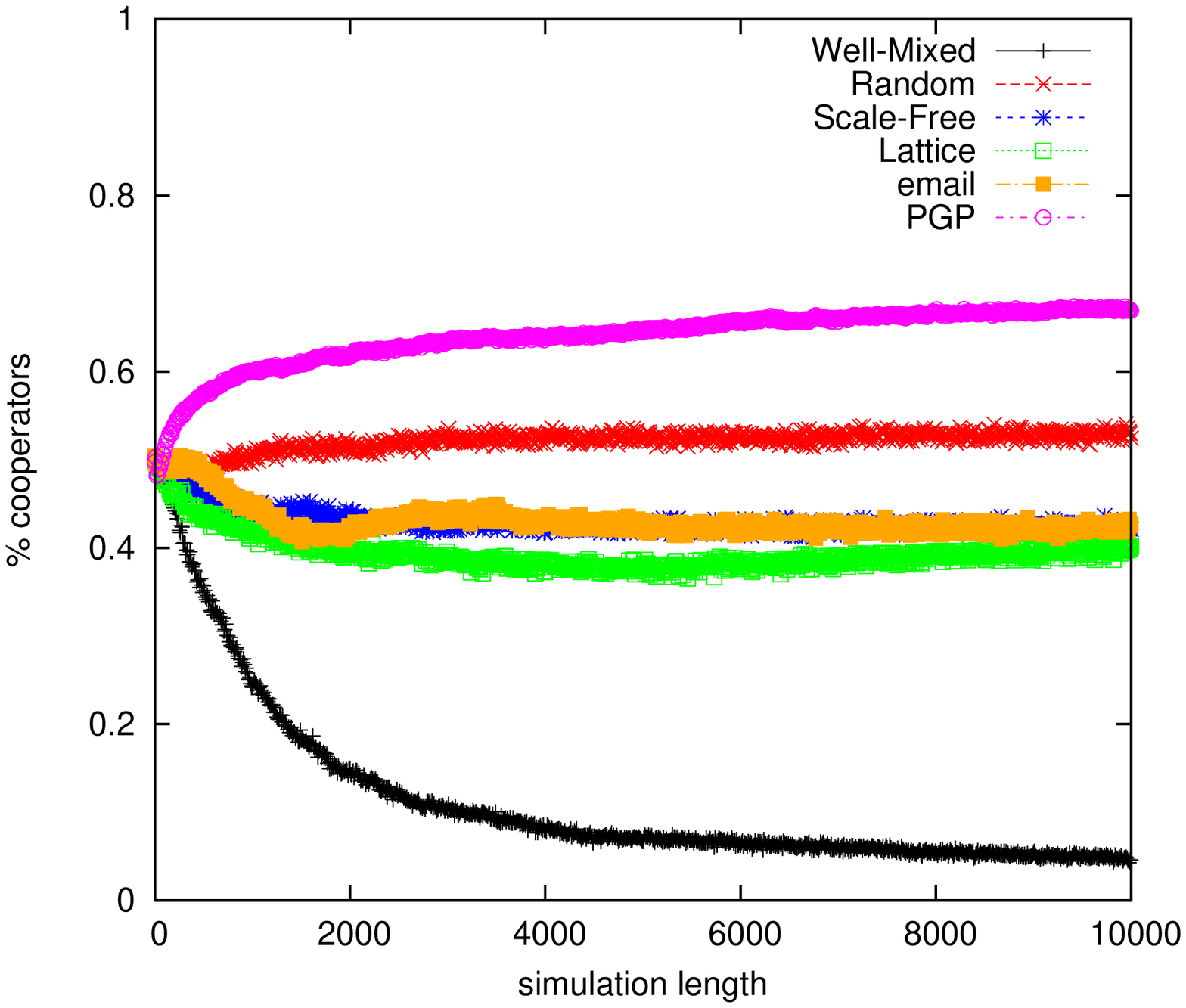}
\newline
\includegraphics[width=0.3\textwidth]{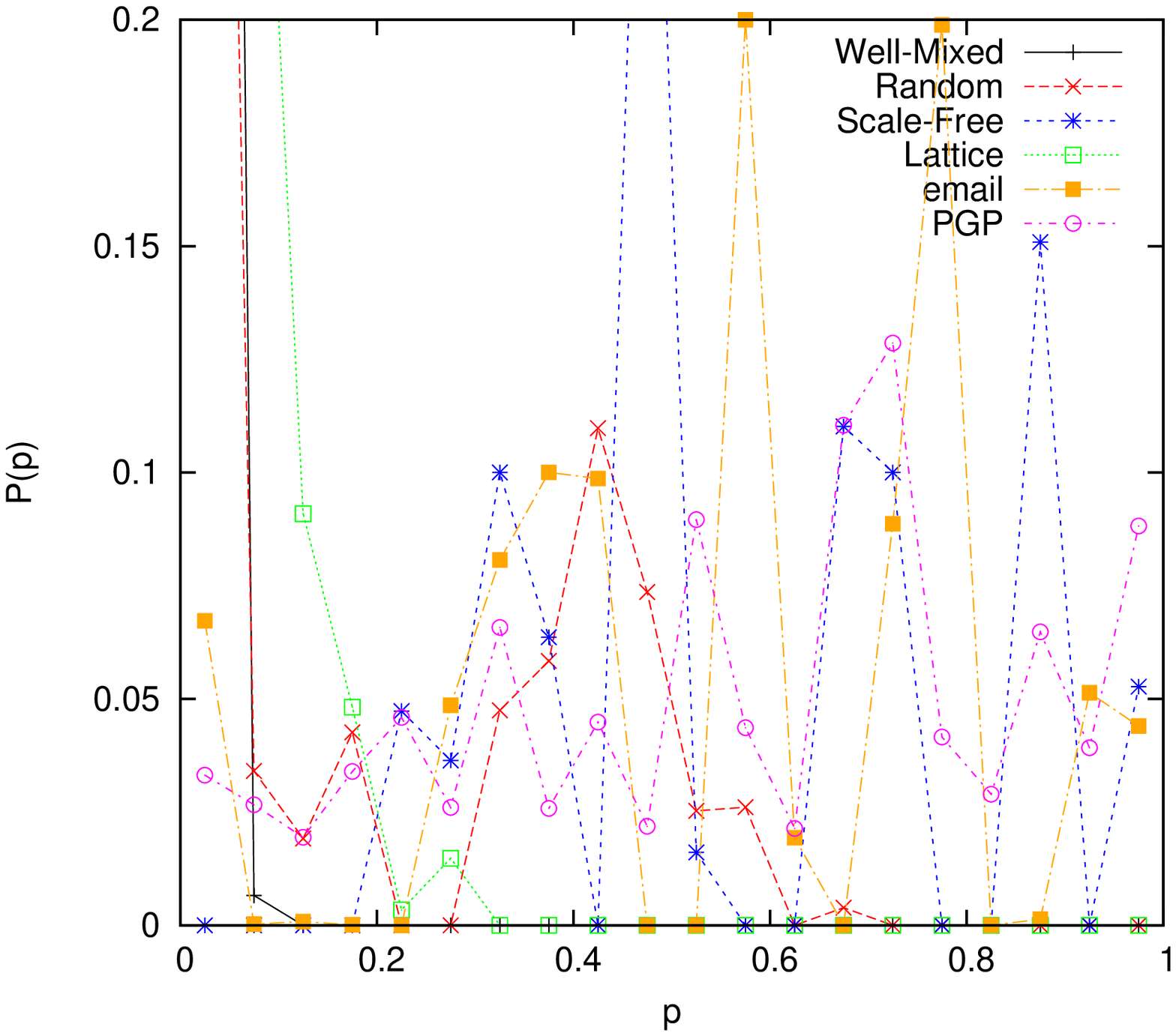}
\hspace{0.5cm}
\includegraphics[width=0.3\textwidth]{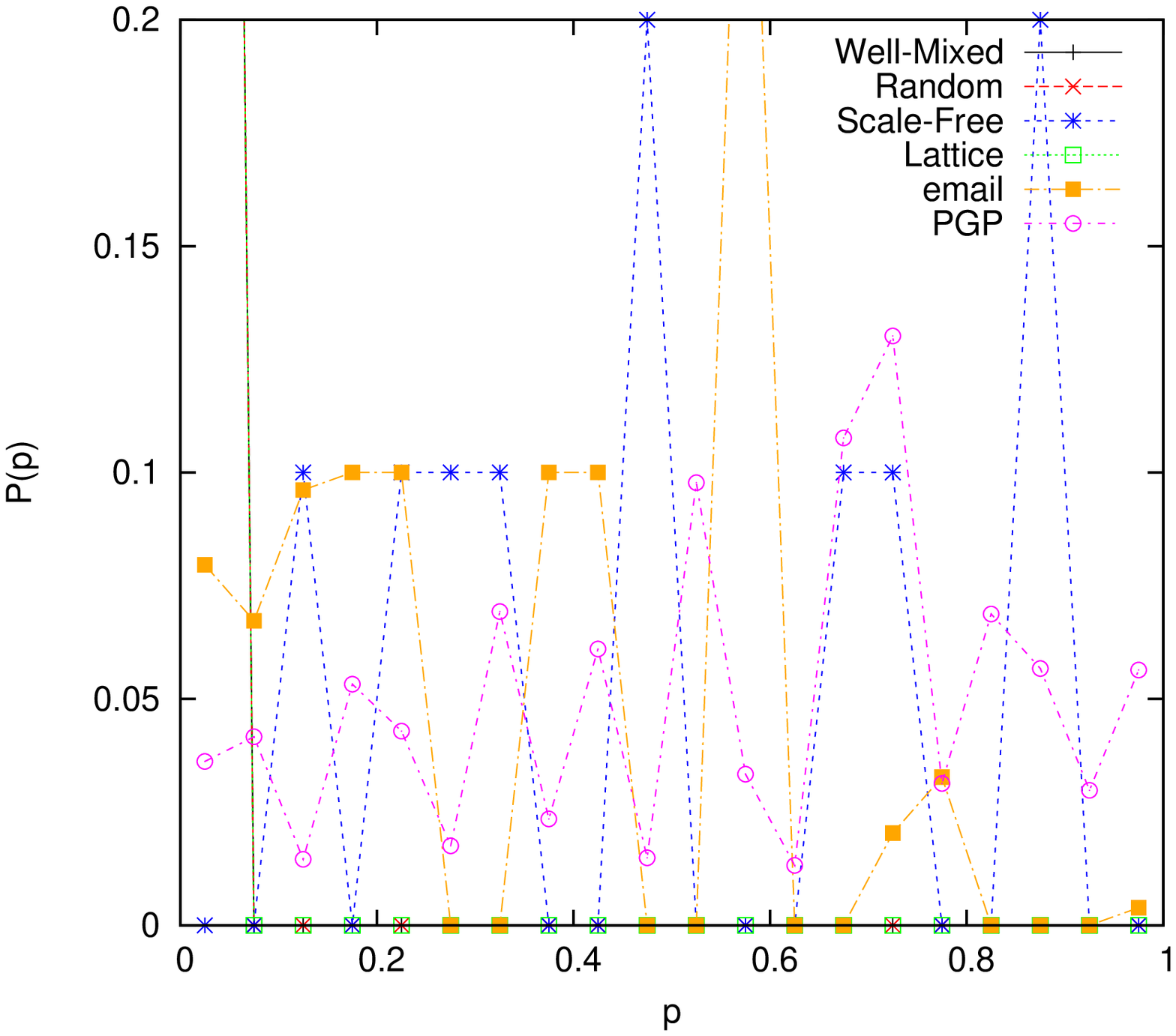}
\hspace{0.5cm}
\includegraphics[width=0.3\textwidth]{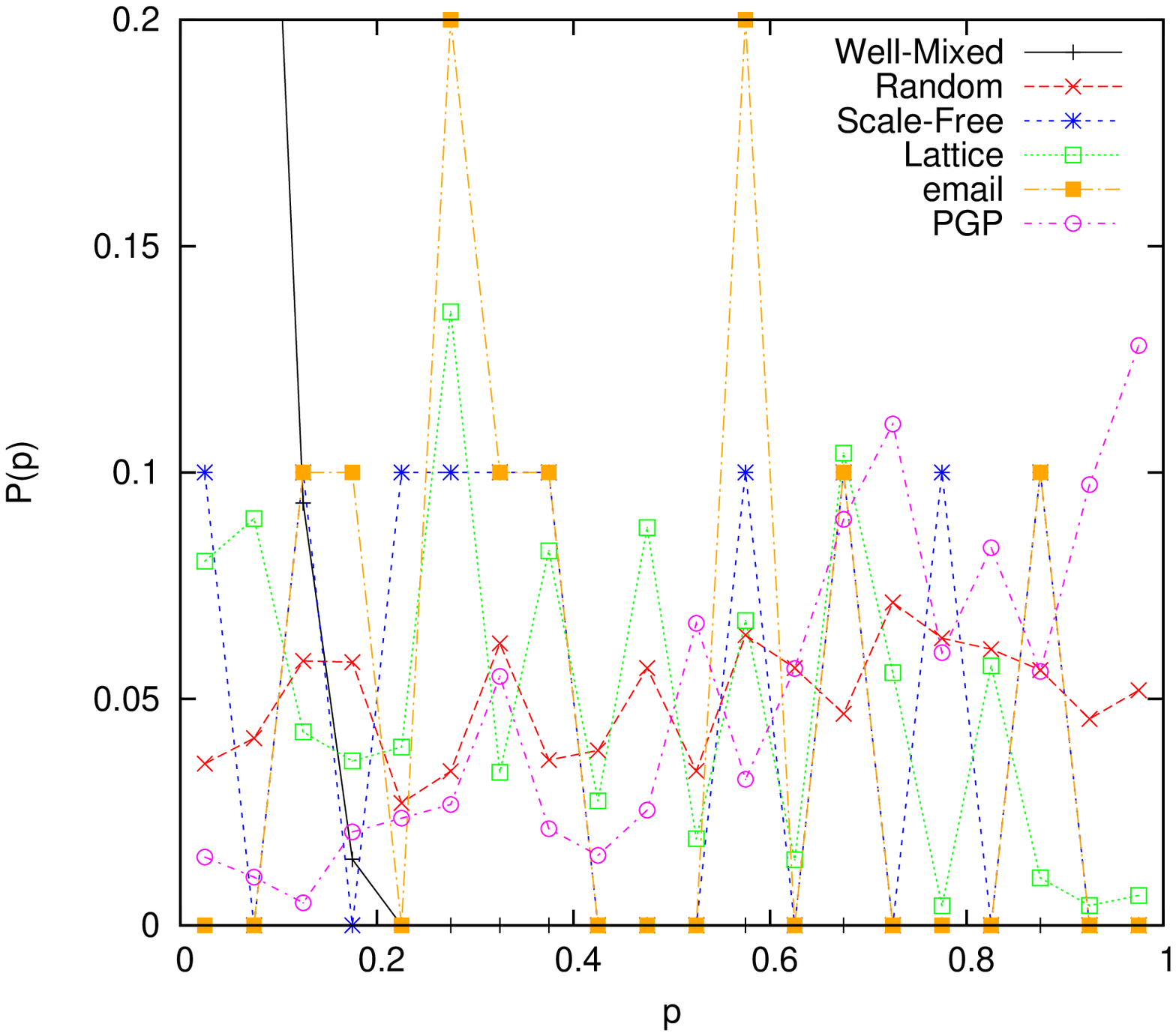}
\caption{Evolution of the level of cooperation $c$ and stationary distribution of $p$ when the evolutionary dynamics is, from left to right: 
Proportional Imitation, Fermi Rule with $\beta=1/2$, Death-Birth rule. Top plots refer to $S=-1/2$, bottom plots to $S=0$. 
$T=3/2$ in all cases. Results are averaged over 10 independent realizations.}
\label{fig.PI_FR_DB}
\end{figure*}

\begin{figure*}
\includegraphics[width=0.3\textwidth]{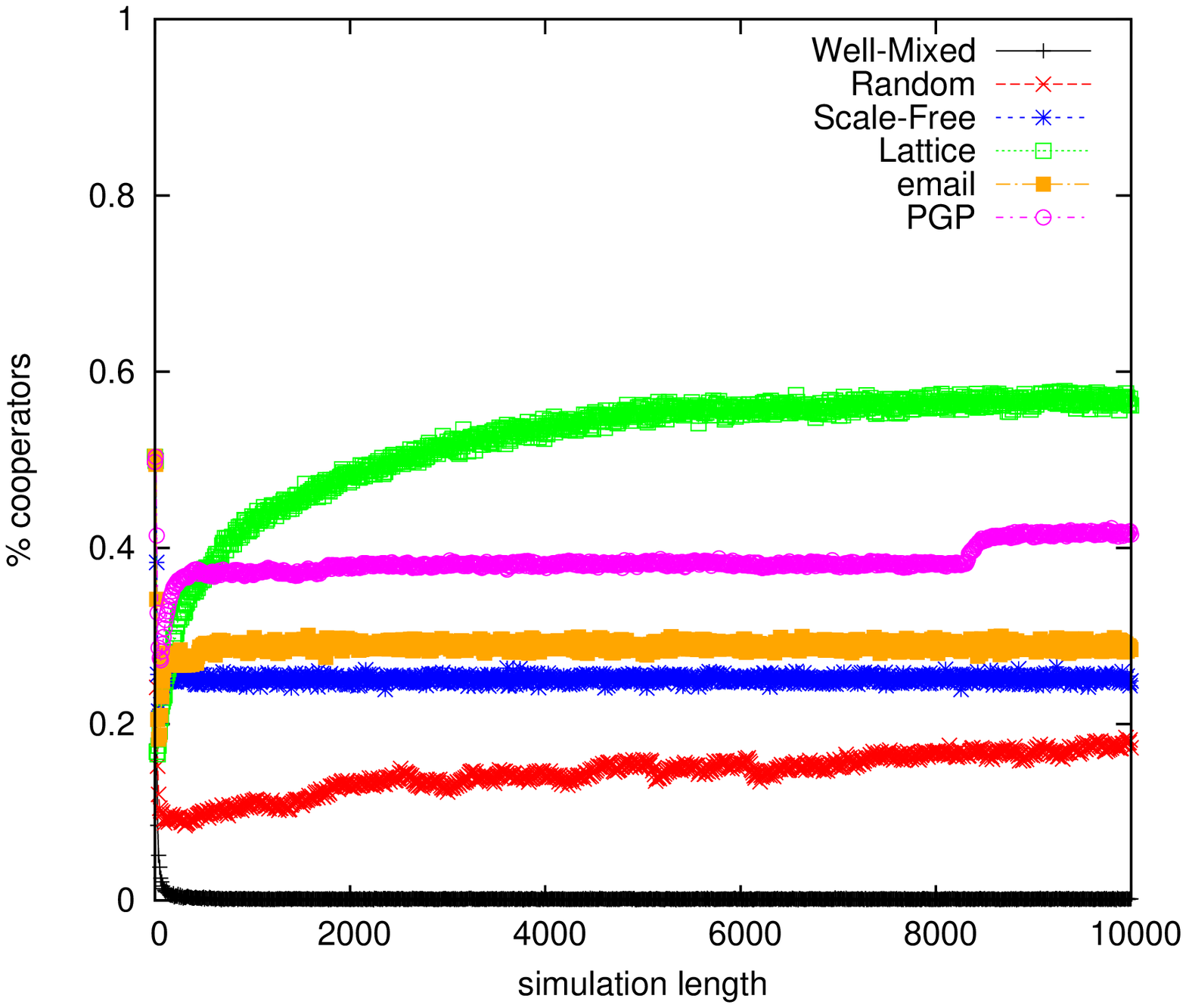}
\hspace{0.5cm}
\includegraphics[width=0.3\textwidth]{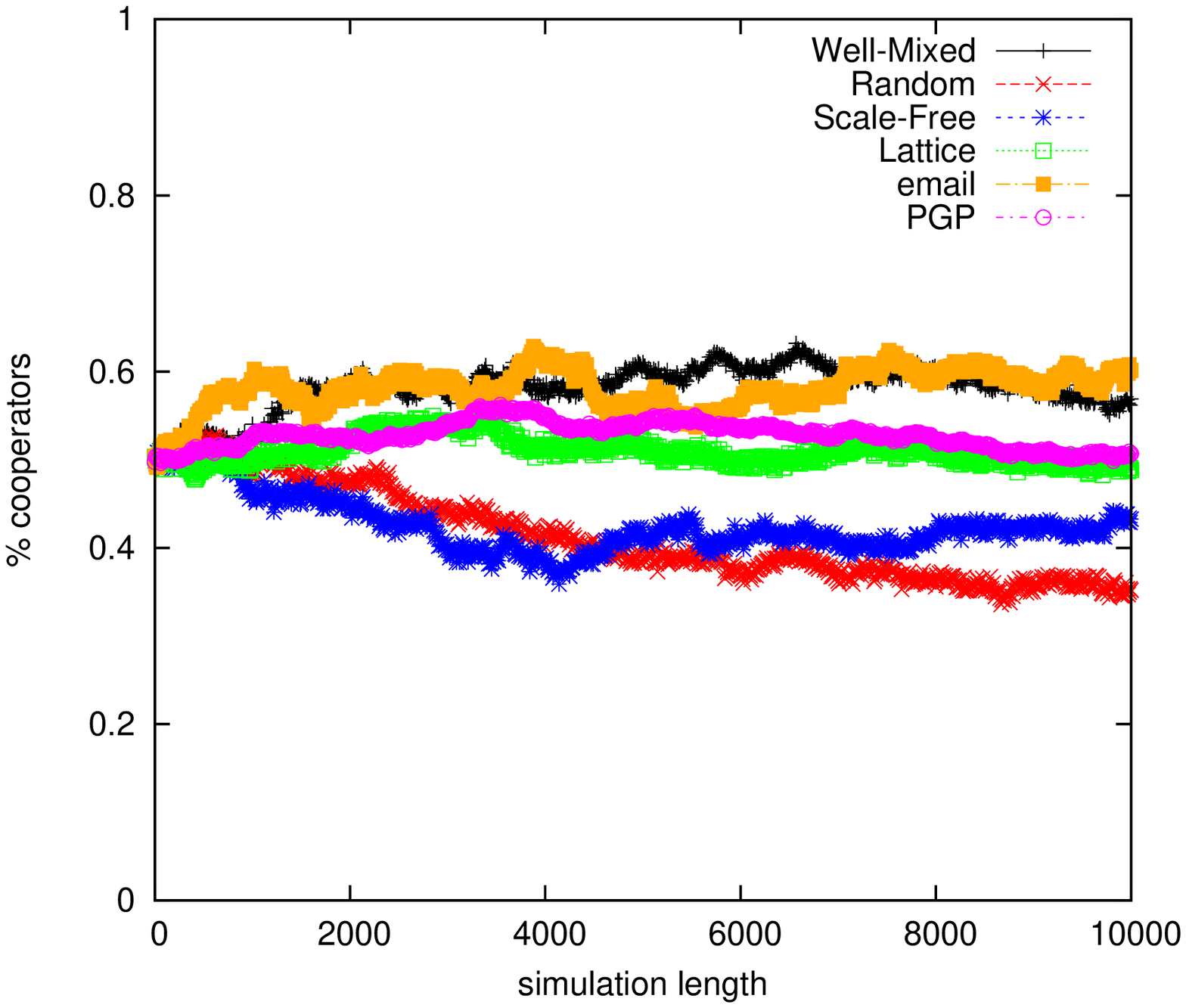}
\hspace{0.5cm}
\includegraphics[width=0.3\textwidth]{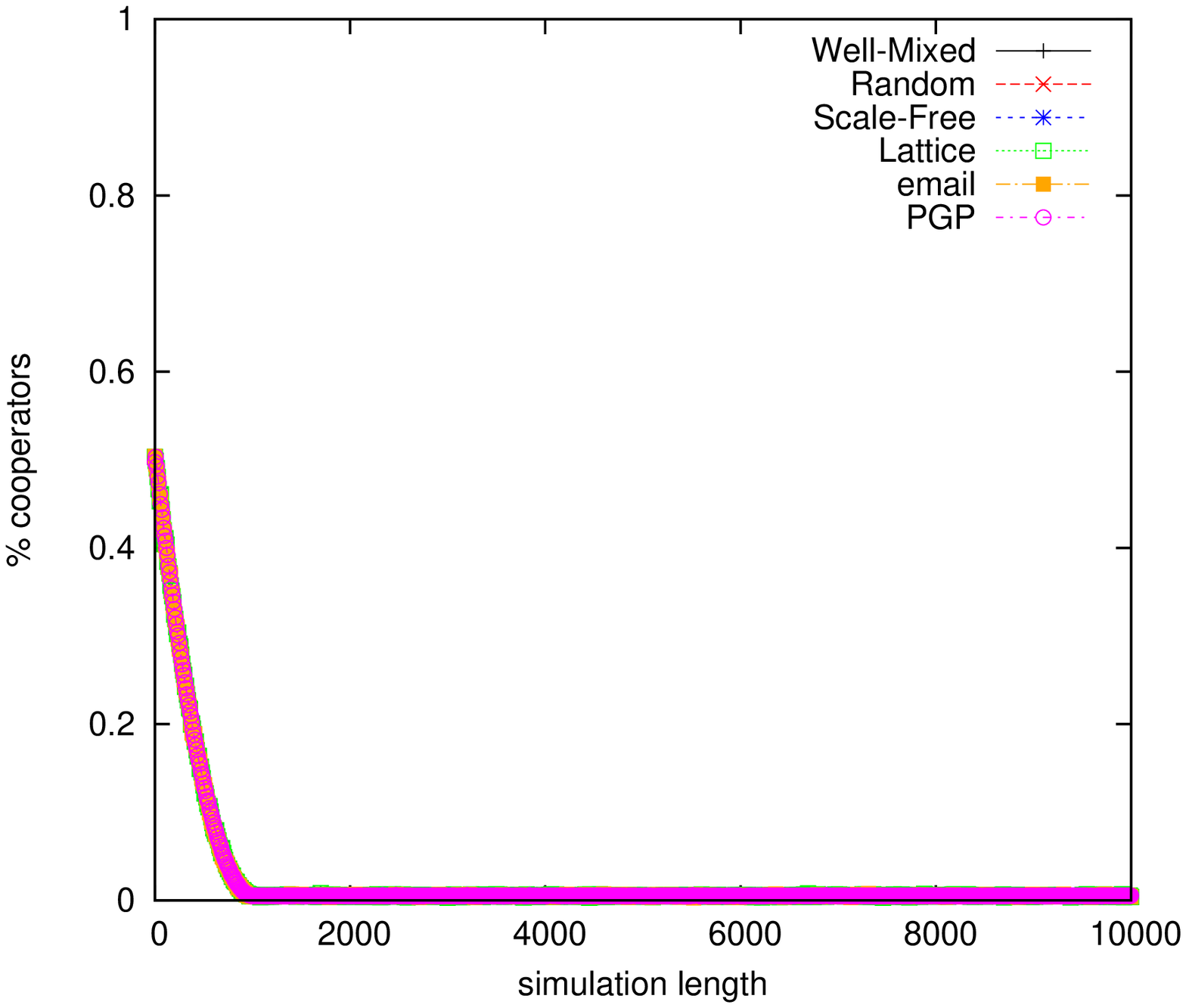}
\newline
\includegraphics[width=0.3\textwidth]{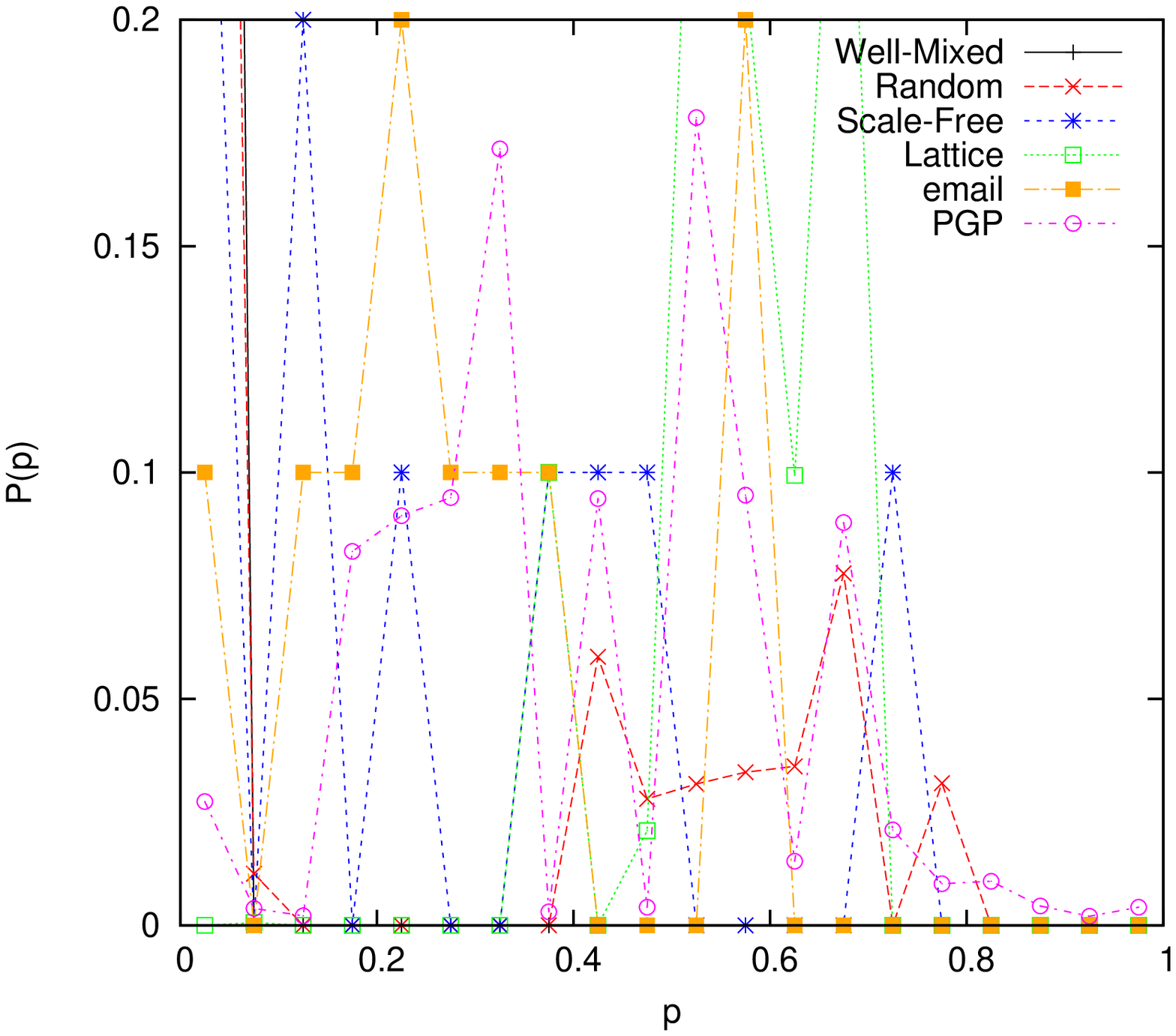}
\hspace{0.5cm}
\includegraphics[width=0.3\textwidth]{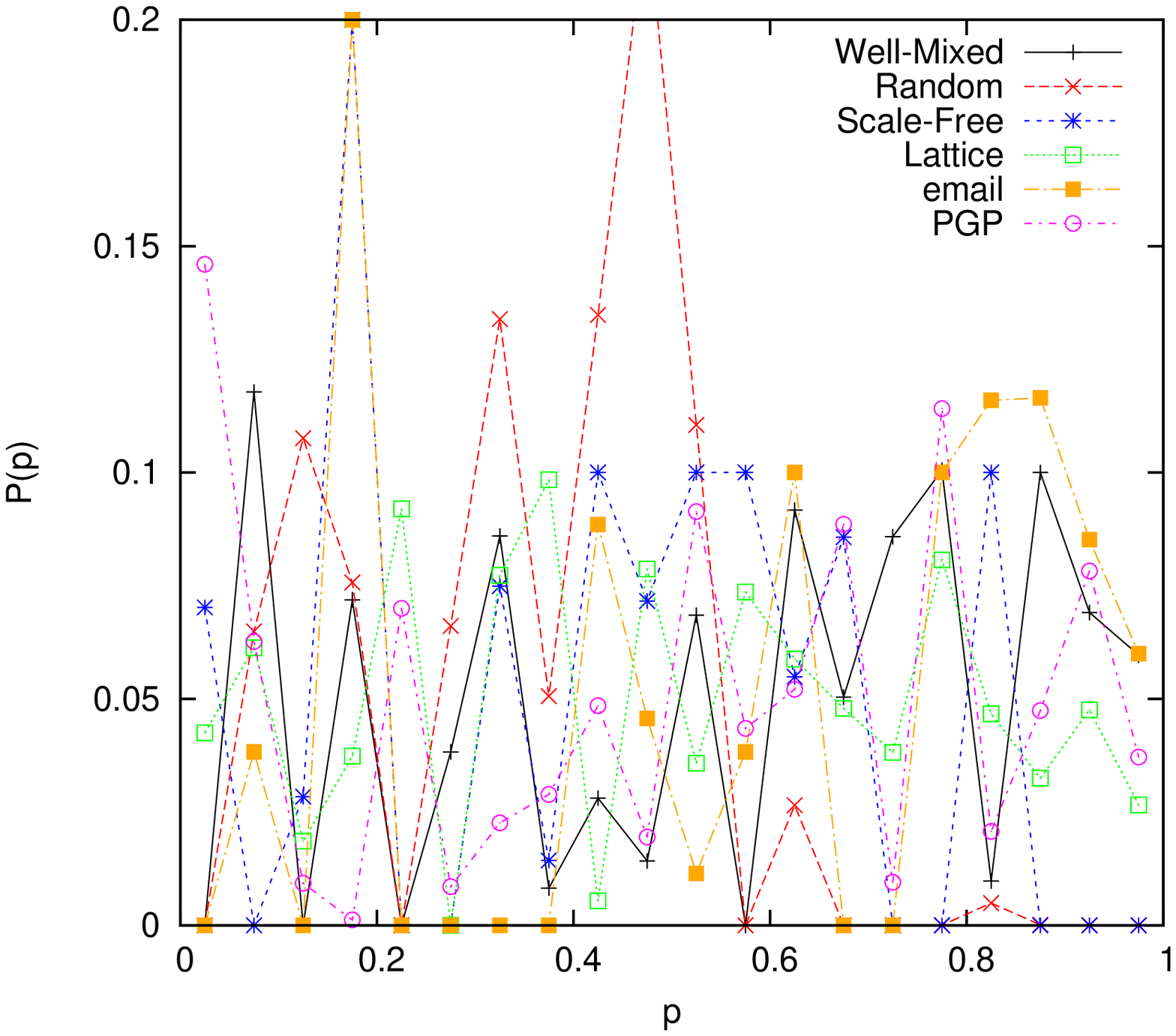}
\hspace{0.5cm}
\includegraphics[width=0.3\textwidth]{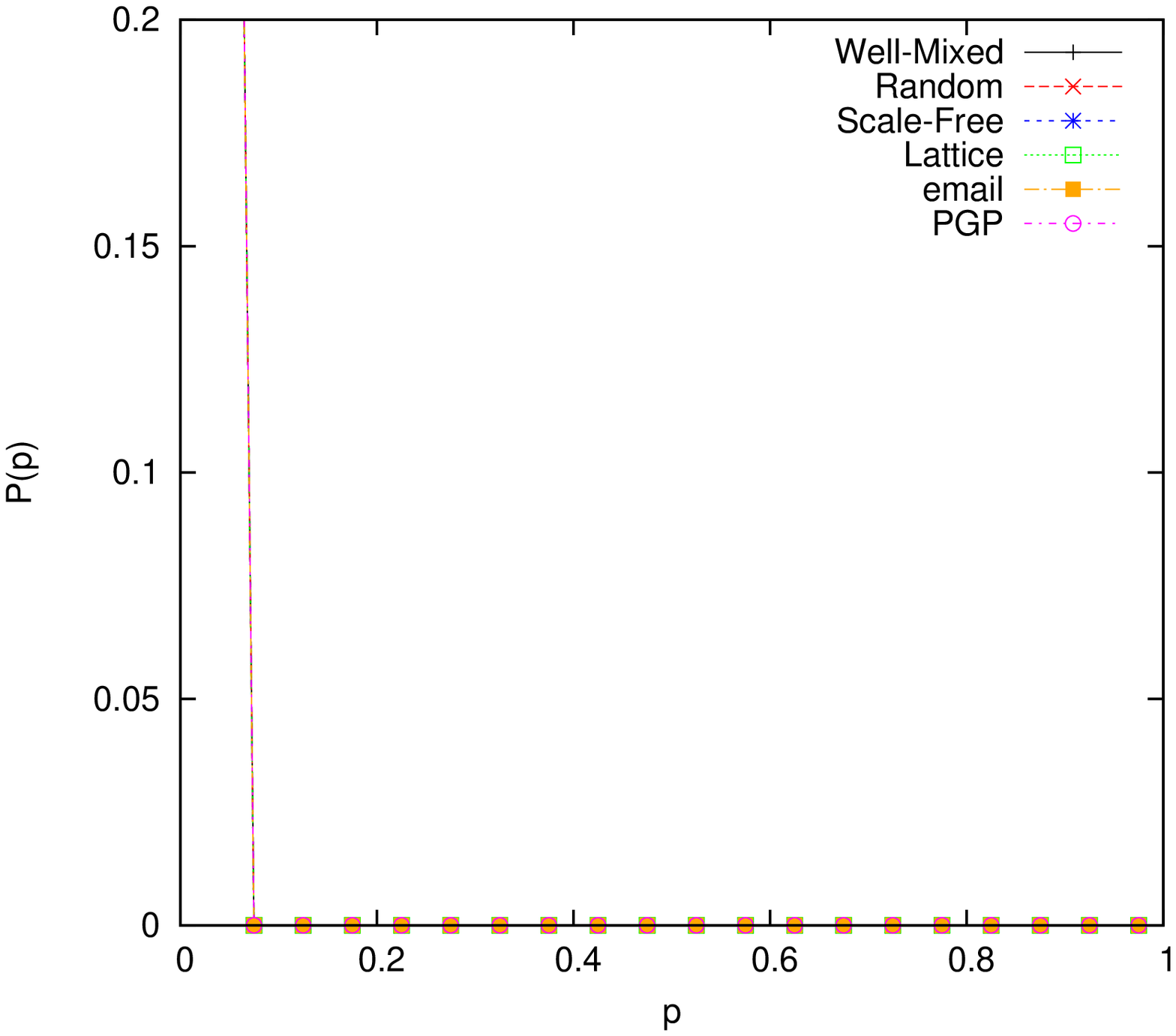}
\newline
\vspace{0.5cm}
\newline
\includegraphics[width=0.3\textwidth]{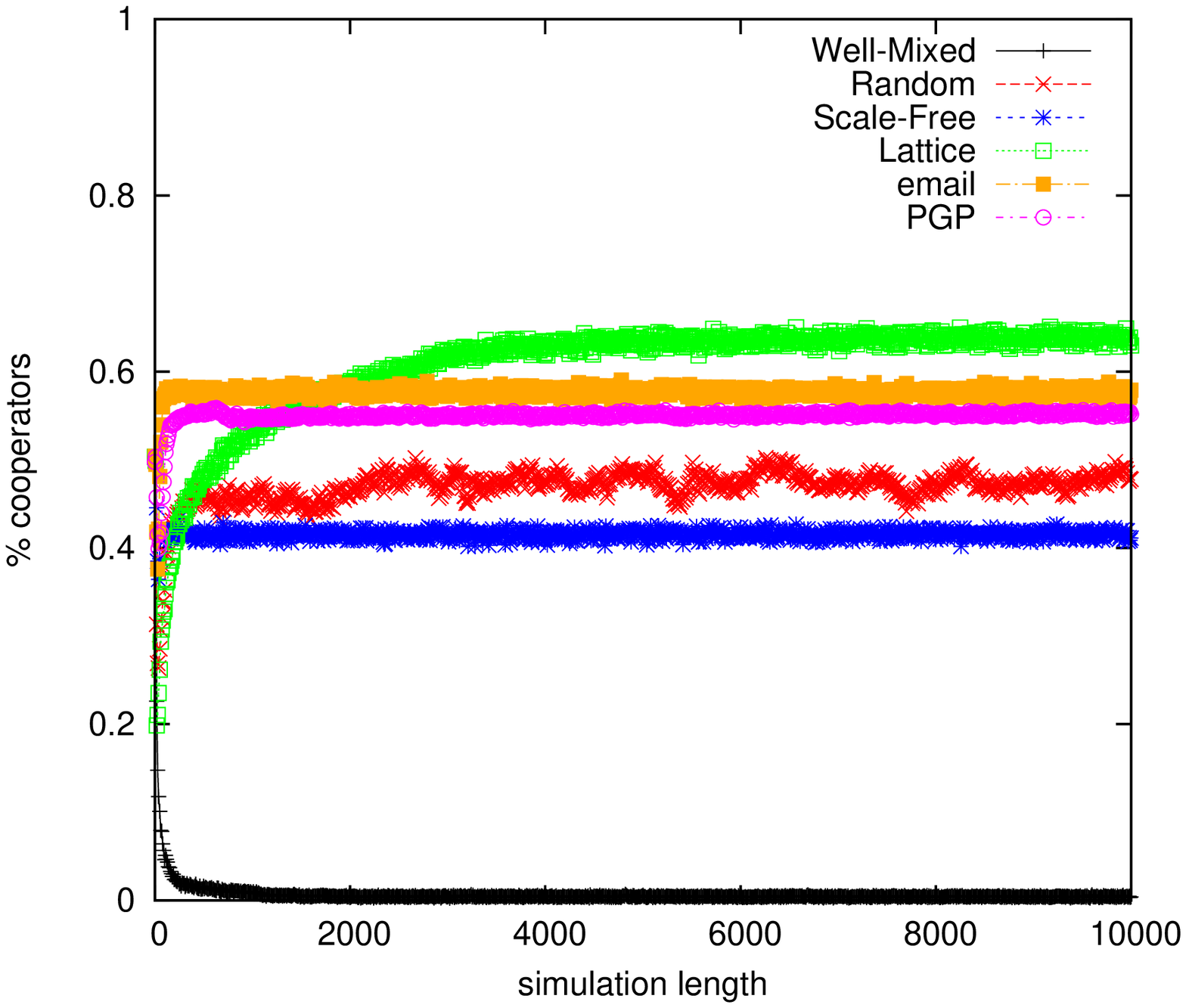}
\hspace{0.5cm}
\includegraphics[width=0.3\textwidth]{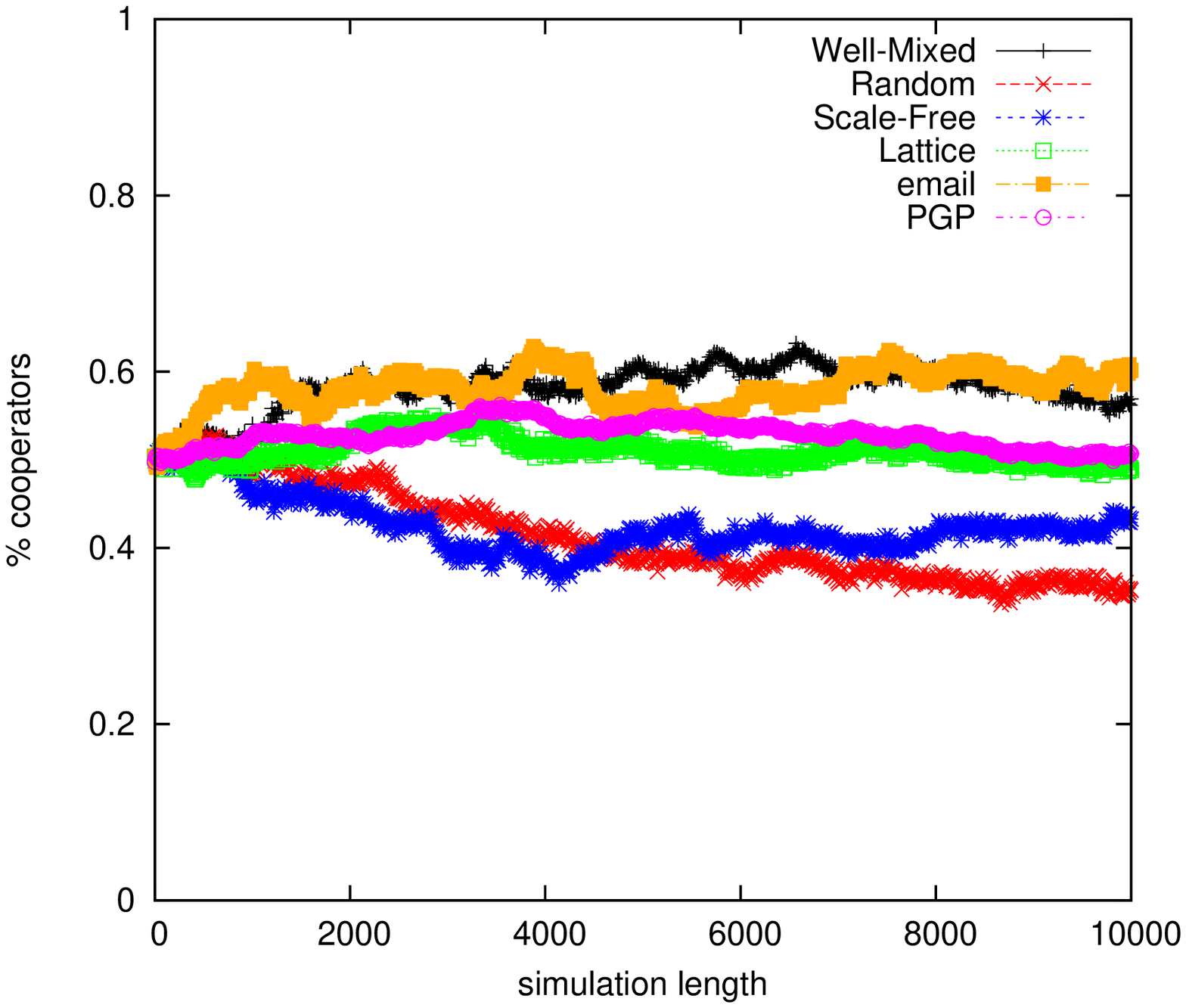}
\hspace{0.5cm}
\includegraphics[width=0.3\textwidth]{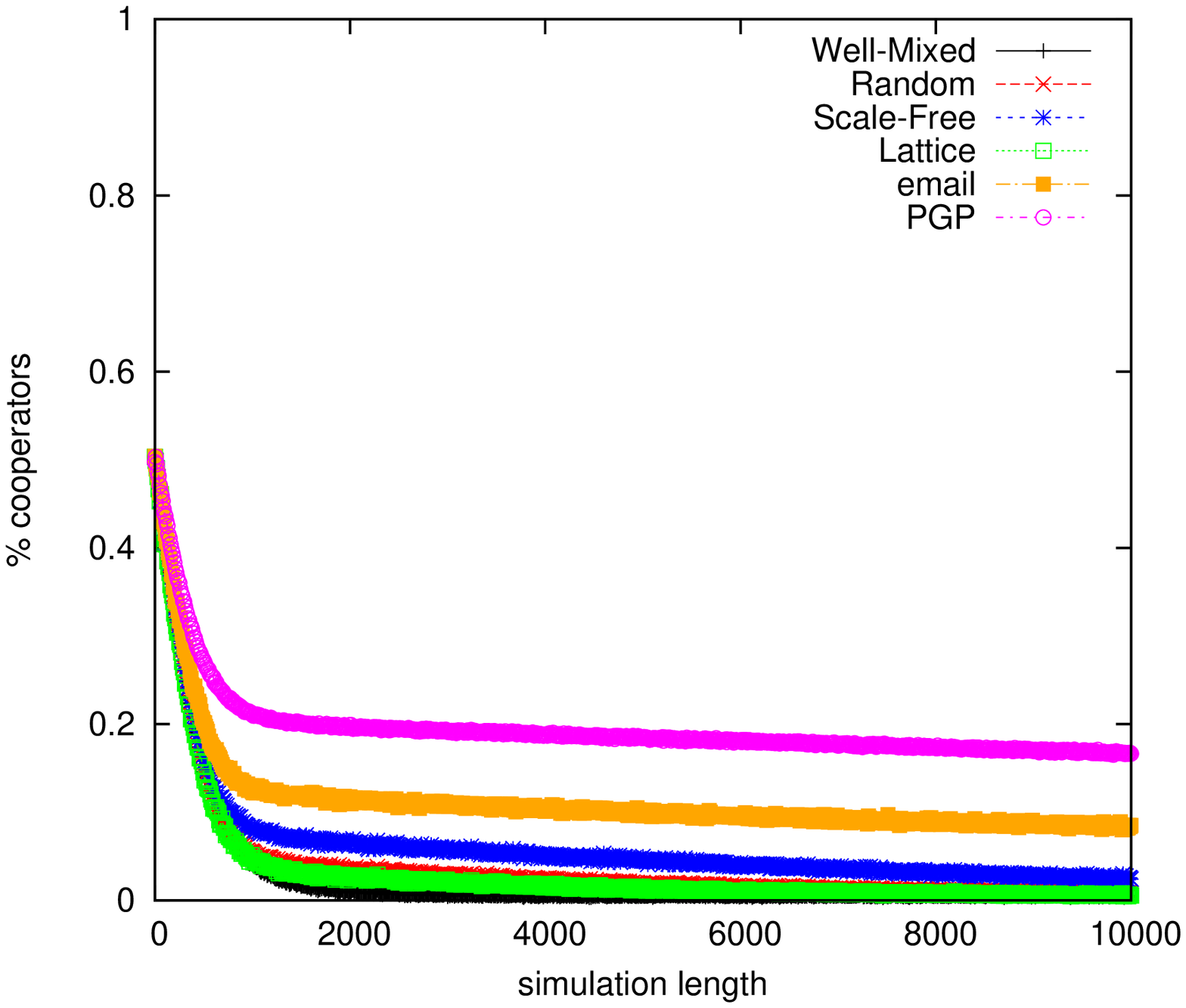}
\newline
\includegraphics[width=0.3\textwidth]{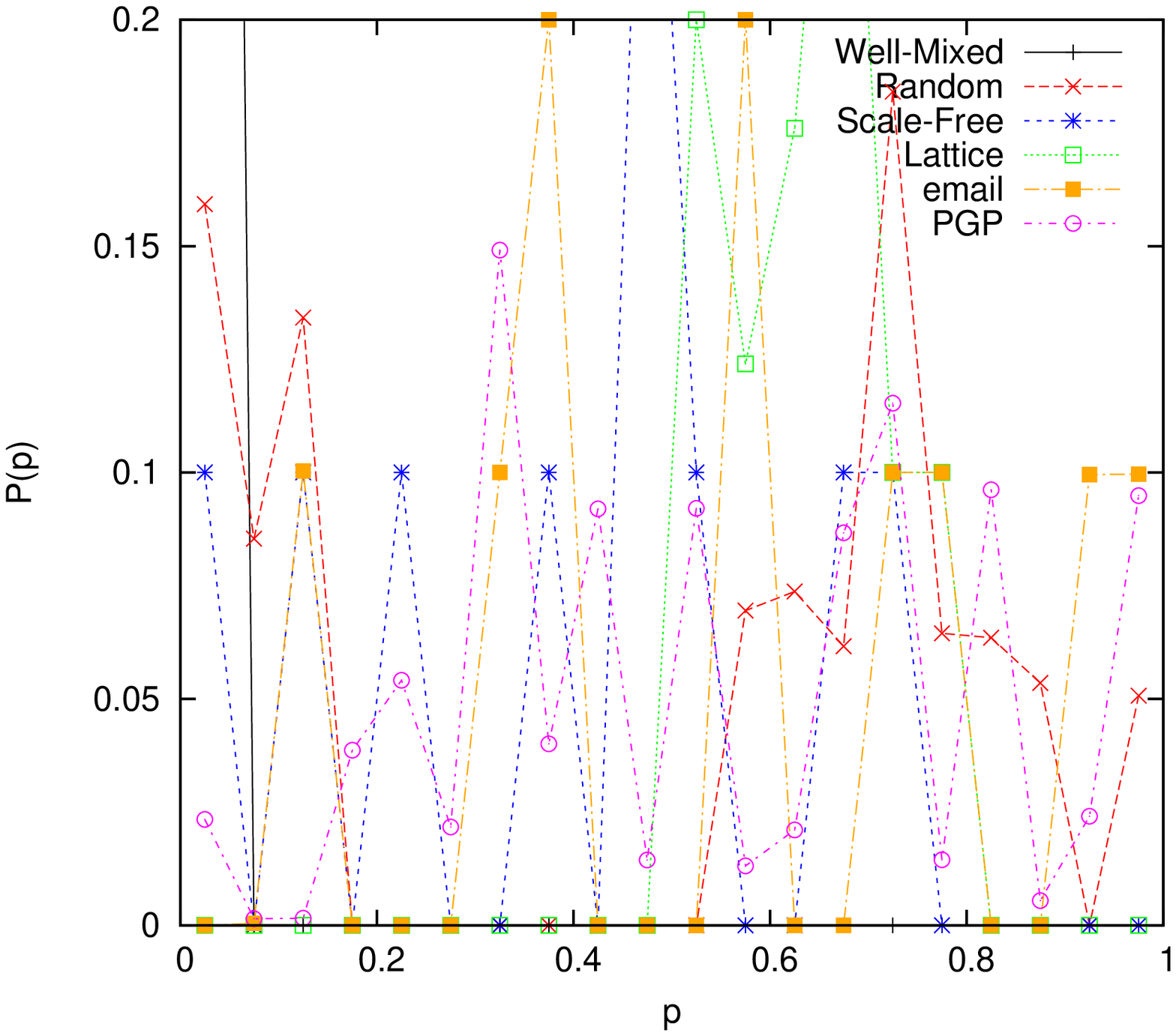}
\hspace{0.5cm}
\includegraphics[width=0.3\textwidth]{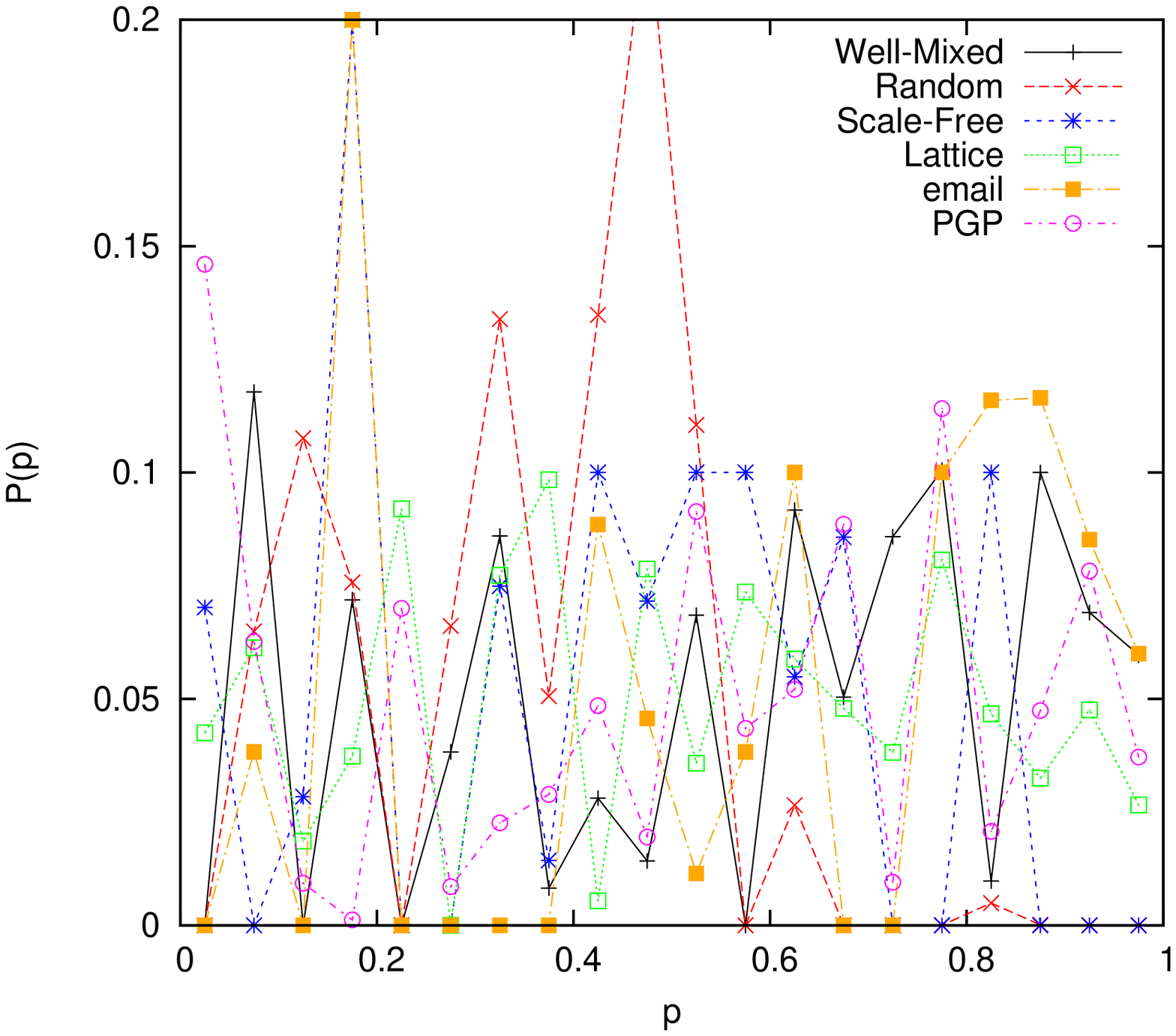}
\hspace{0.5cm}
\includegraphics[width=0.3\textwidth]{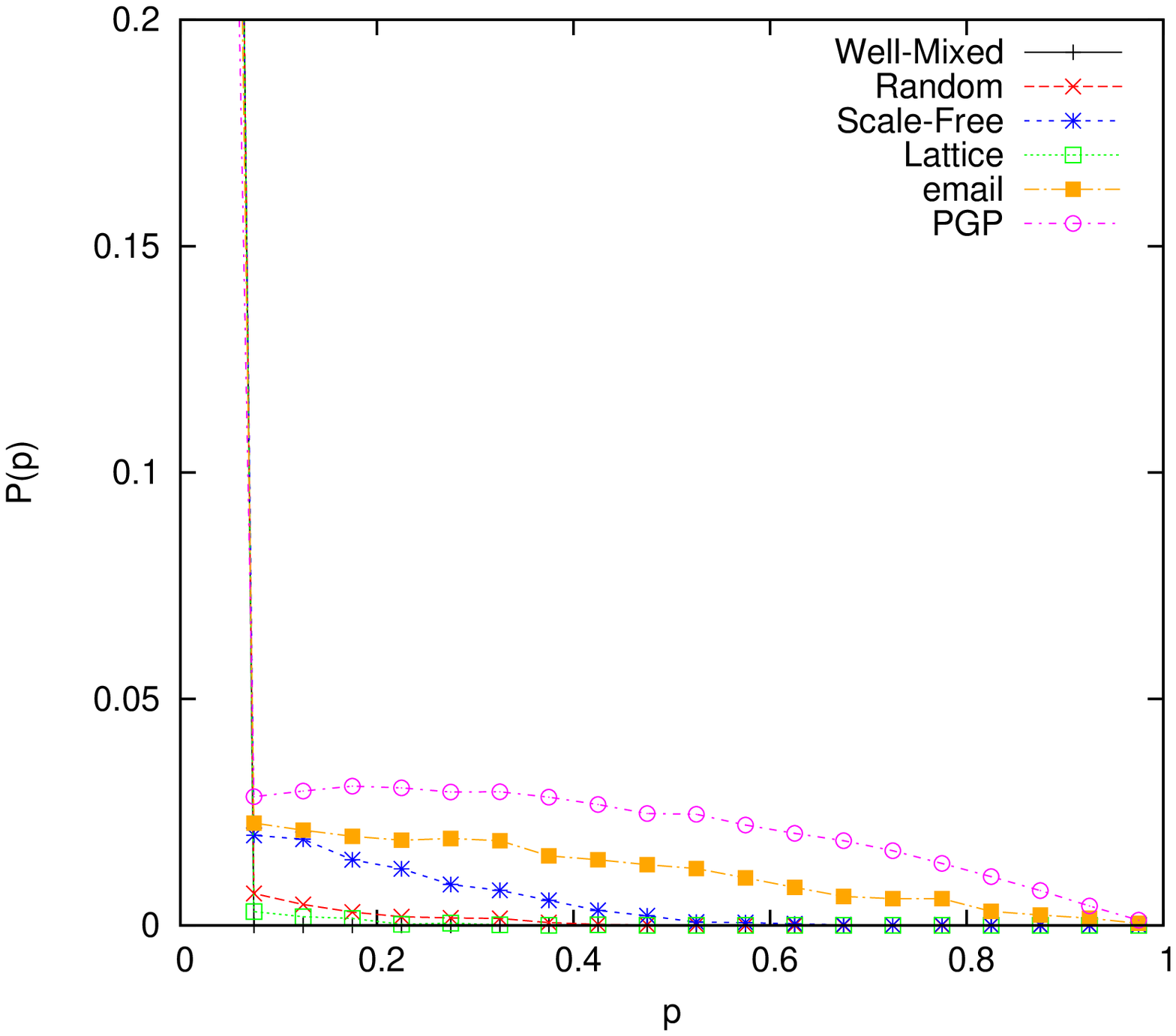}
\caption{Evolution of the level of cooperation $c$ and stationary distribution of $p$ when the evolutionary dynamics is, from left to right: 
Unconditional Imitation, Voter model, Best Response with $\delta=10^{-2}$. Top plots refer to $S=-1/2$, bottom plots to $S=0$. 
$T=3/2$ in all cases. Results are averaged over 10 independent realizations.}
\label{fig.UI_VM_BR}
\end{figure*}

\begin{figure*}
\includegraphics[width=0.3\textwidth]{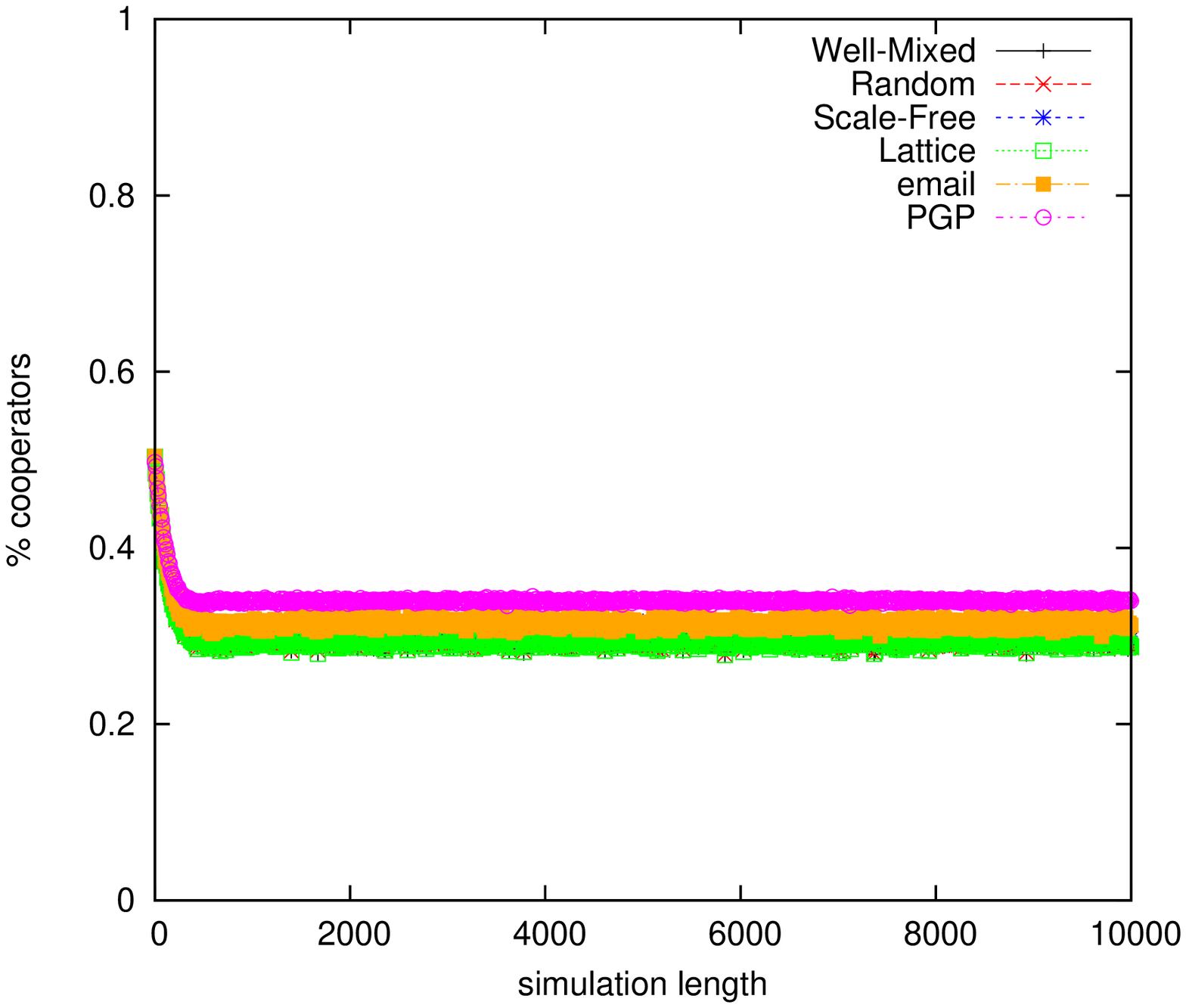}
\hspace{0.5cm}
\includegraphics[width=0.3\textwidth]{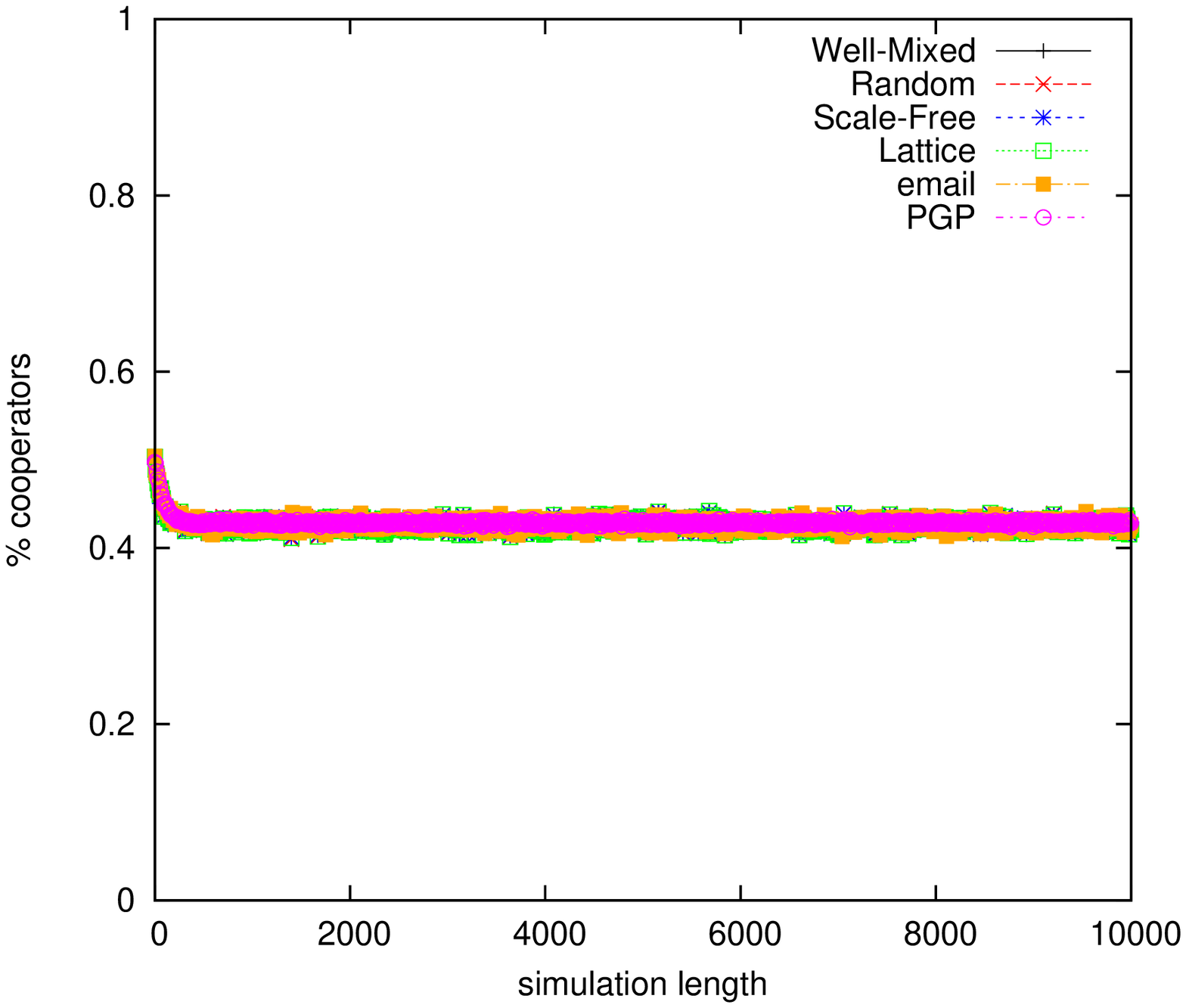}
\hspace{0.5cm}
\includegraphics[width=0.3\textwidth]{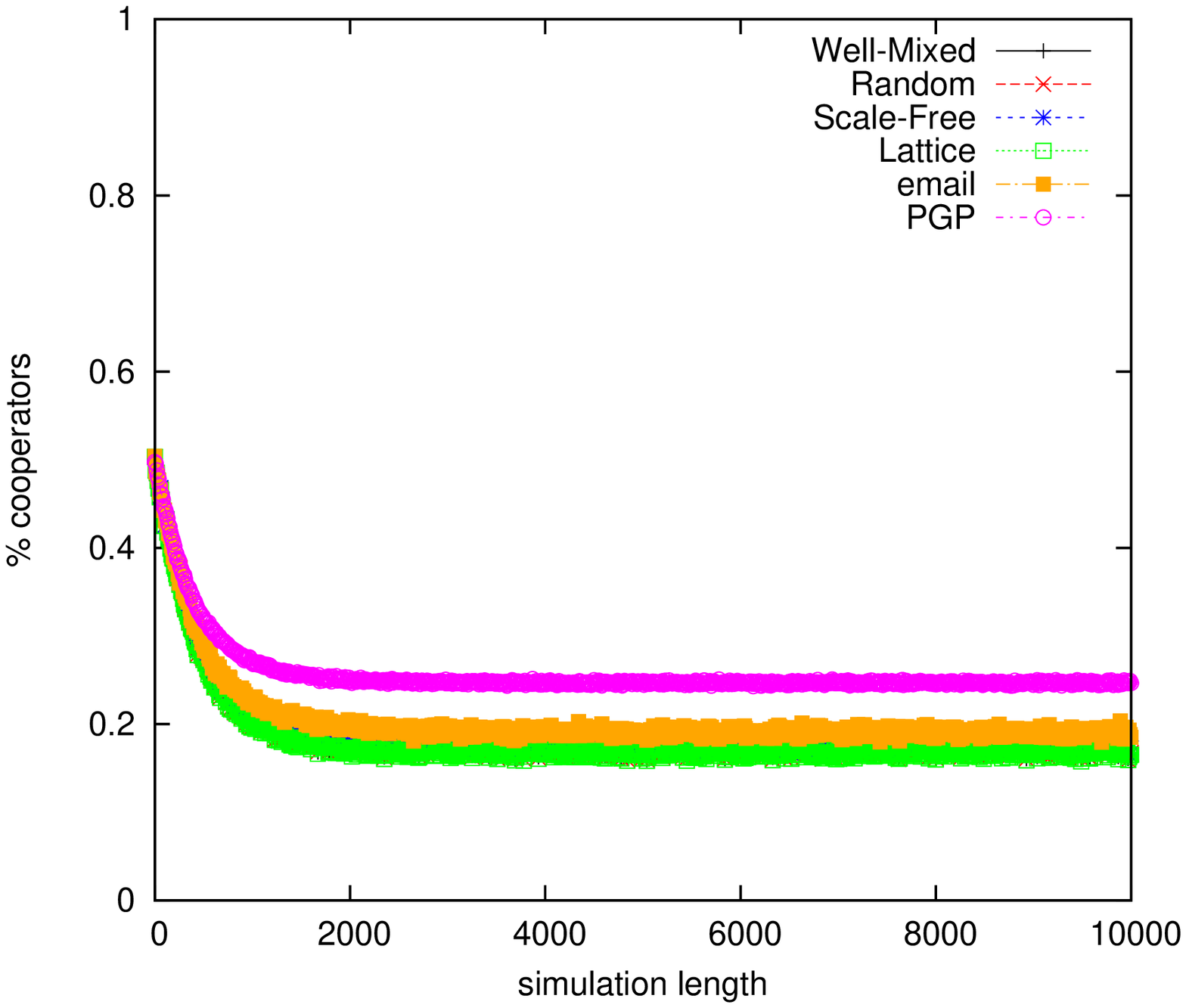}
\newline
\includegraphics[width=0.3\textwidth]{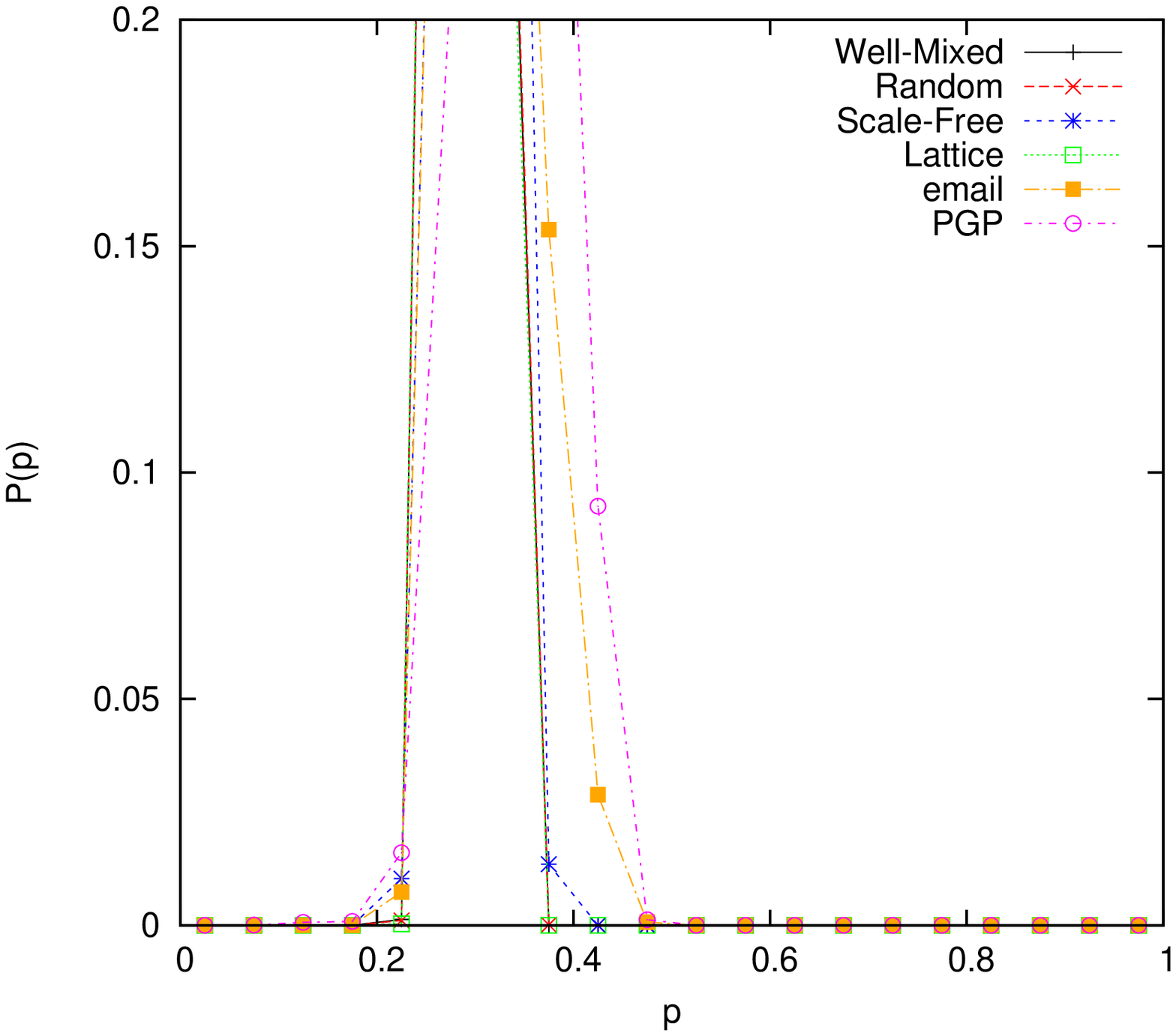}
\hspace{0.5cm}
\includegraphics[width=0.3\textwidth]{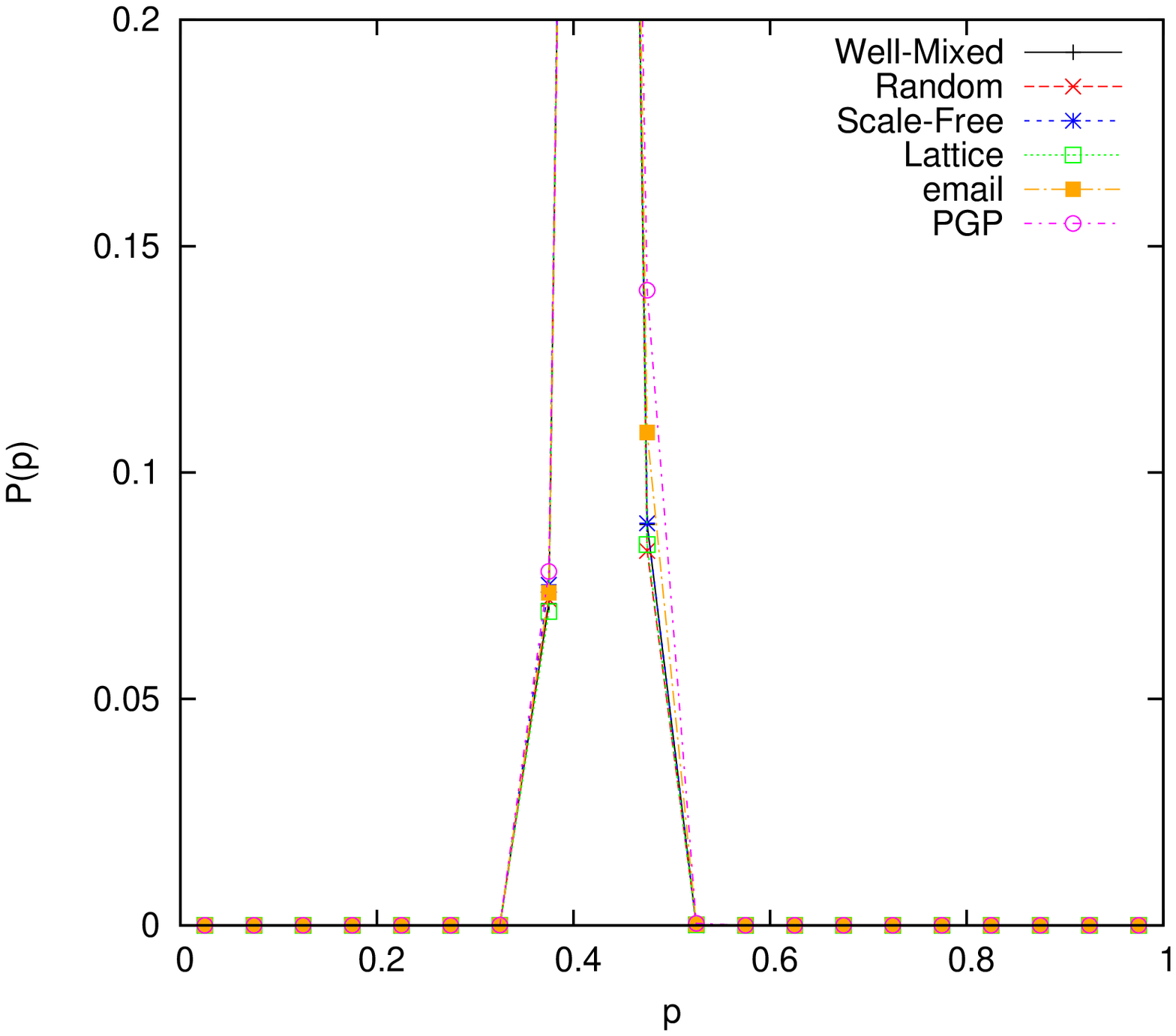}
\hspace{0.5cm}
\includegraphics[width=0.3\textwidth]{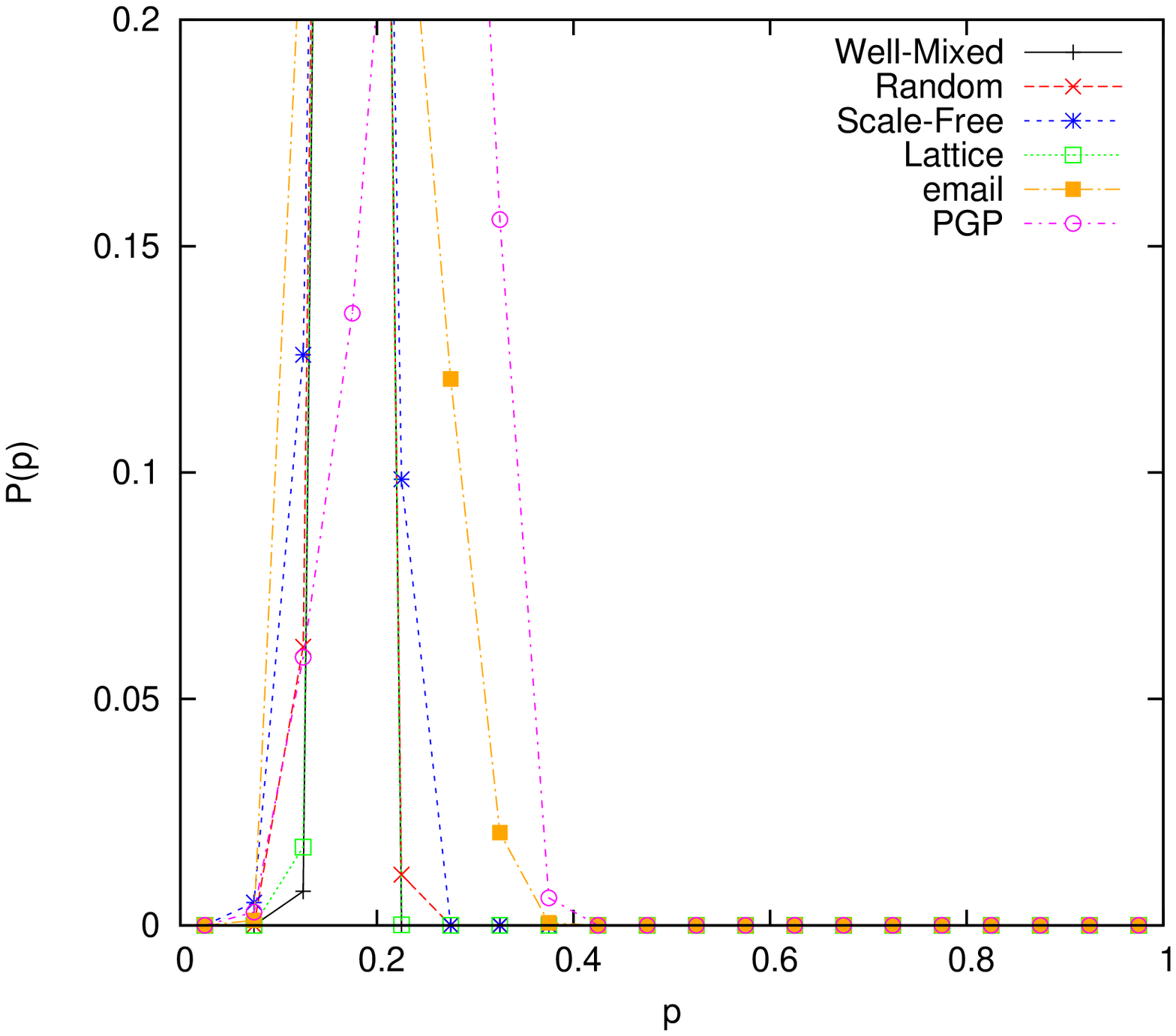}
\newline
\vspace{0.5cm}
\newline
\includegraphics[width=0.3\textwidth]{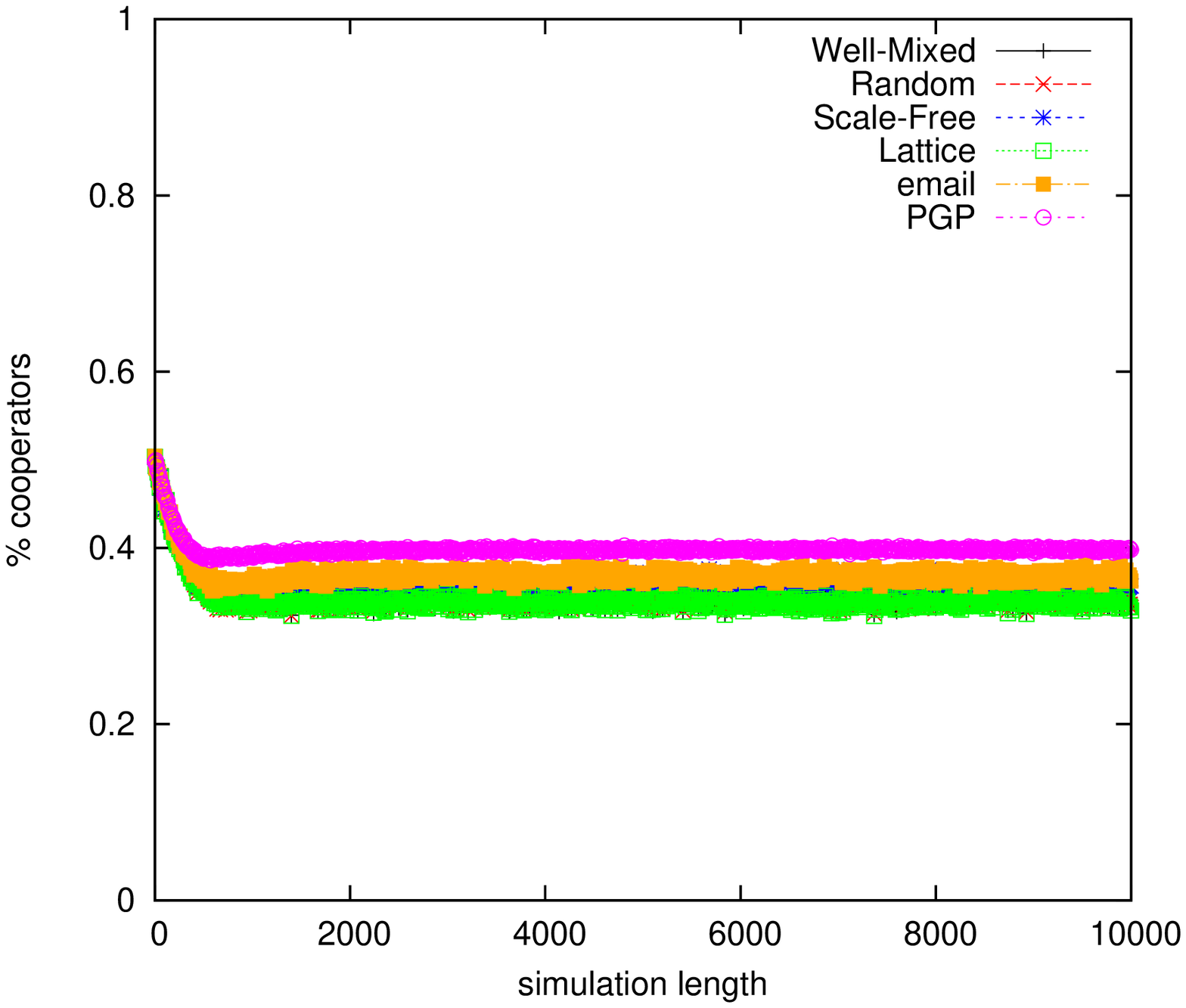}
\hspace{0.5cm}
\includegraphics[width=0.3\textwidth]{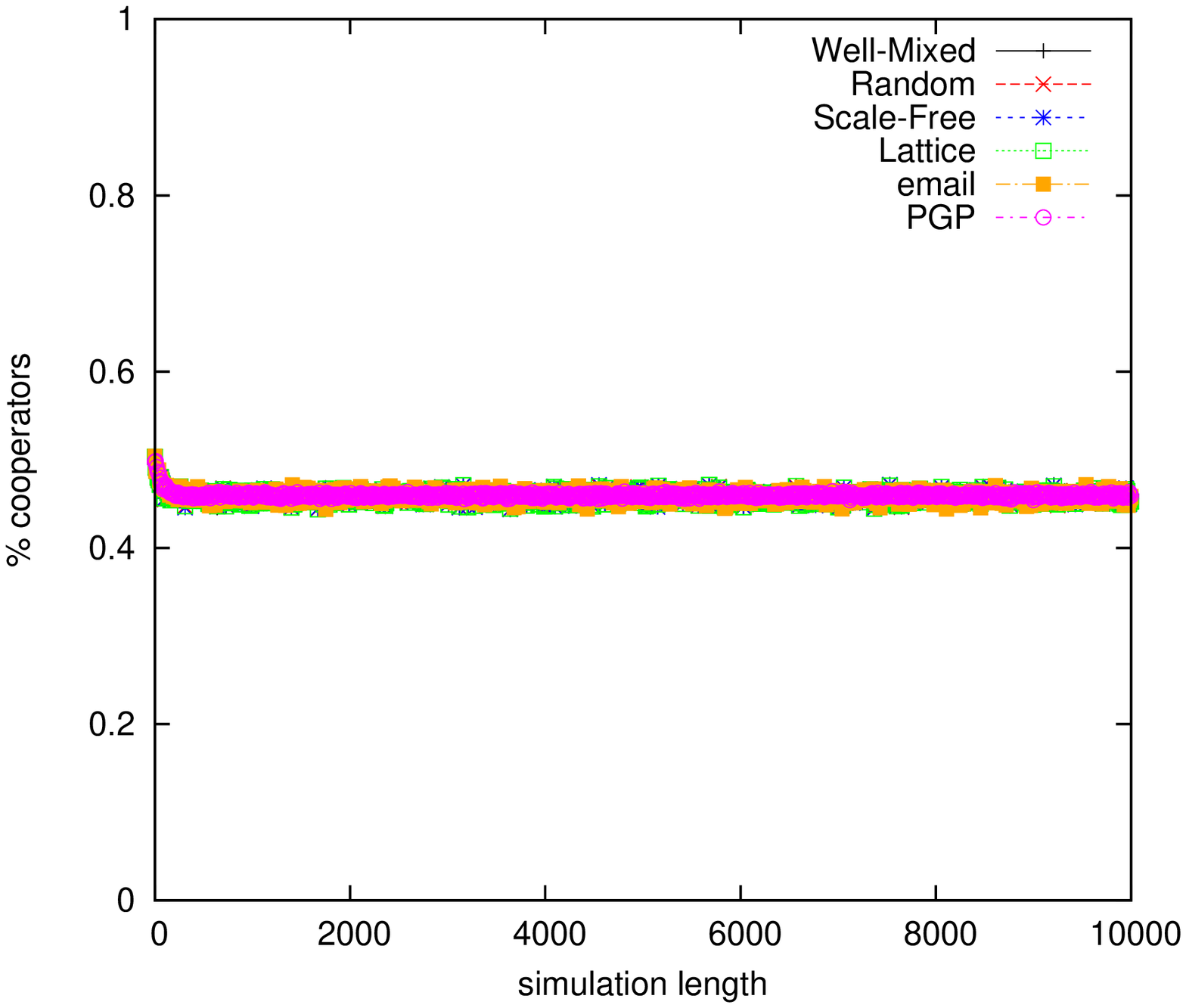}
\hspace{0.5cm}
\includegraphics[width=0.3\textwidth]{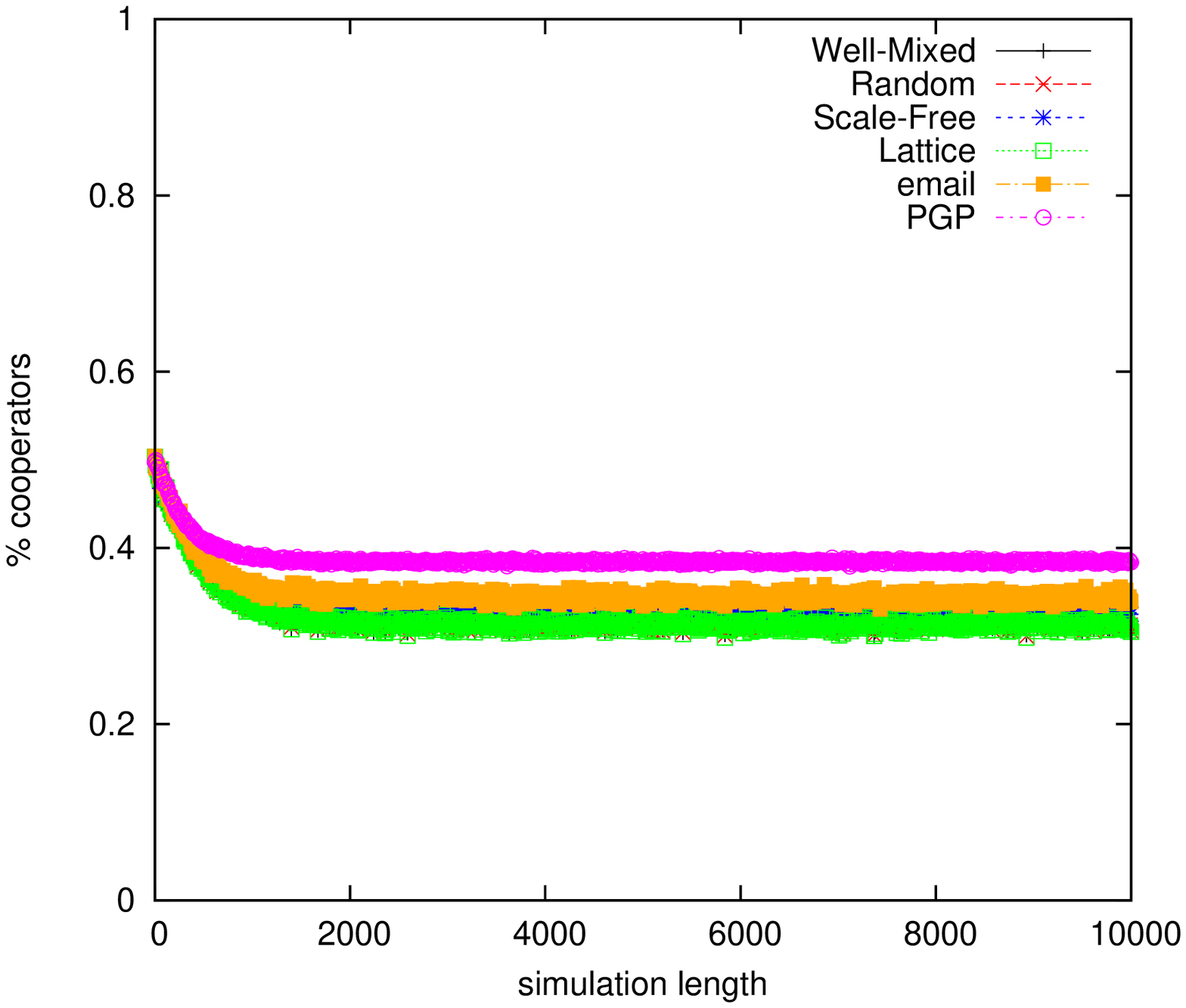}
\newline
\includegraphics[width=0.3\textwidth]{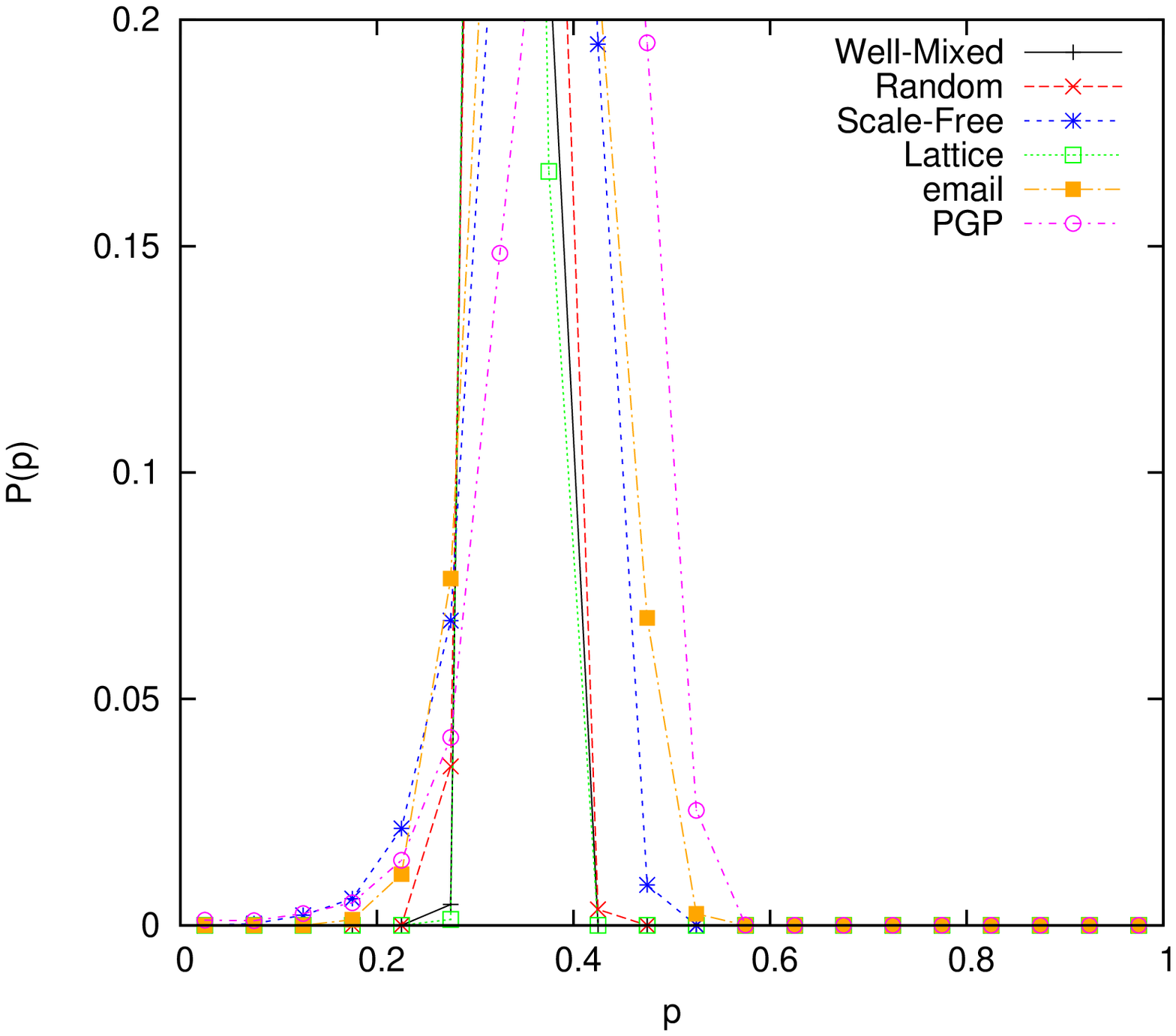}
\hspace{0.5cm}
\includegraphics[width=0.3\textwidth]{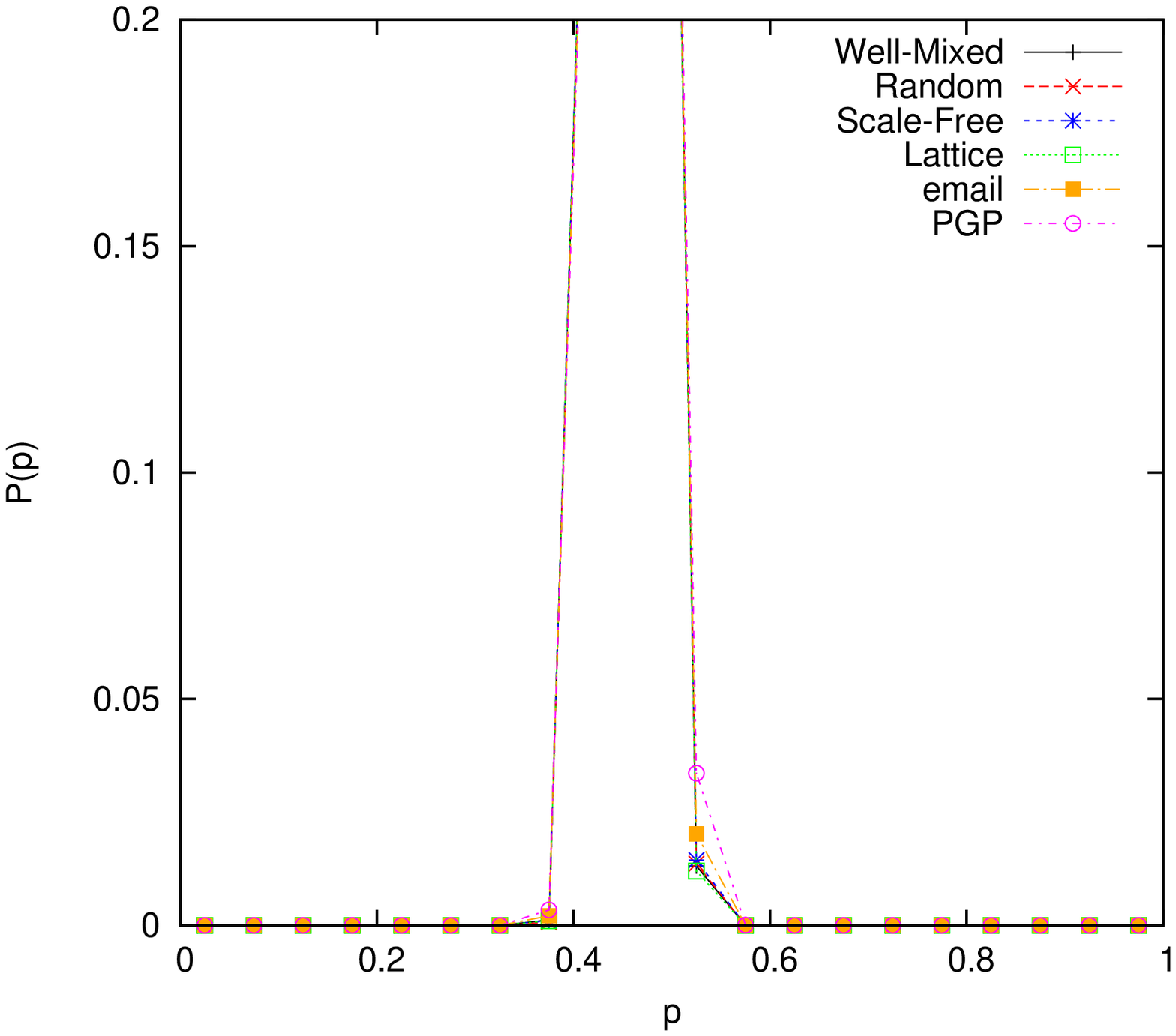}
\hspace{0.5cm}
\includegraphics[width=0.3\textwidth]{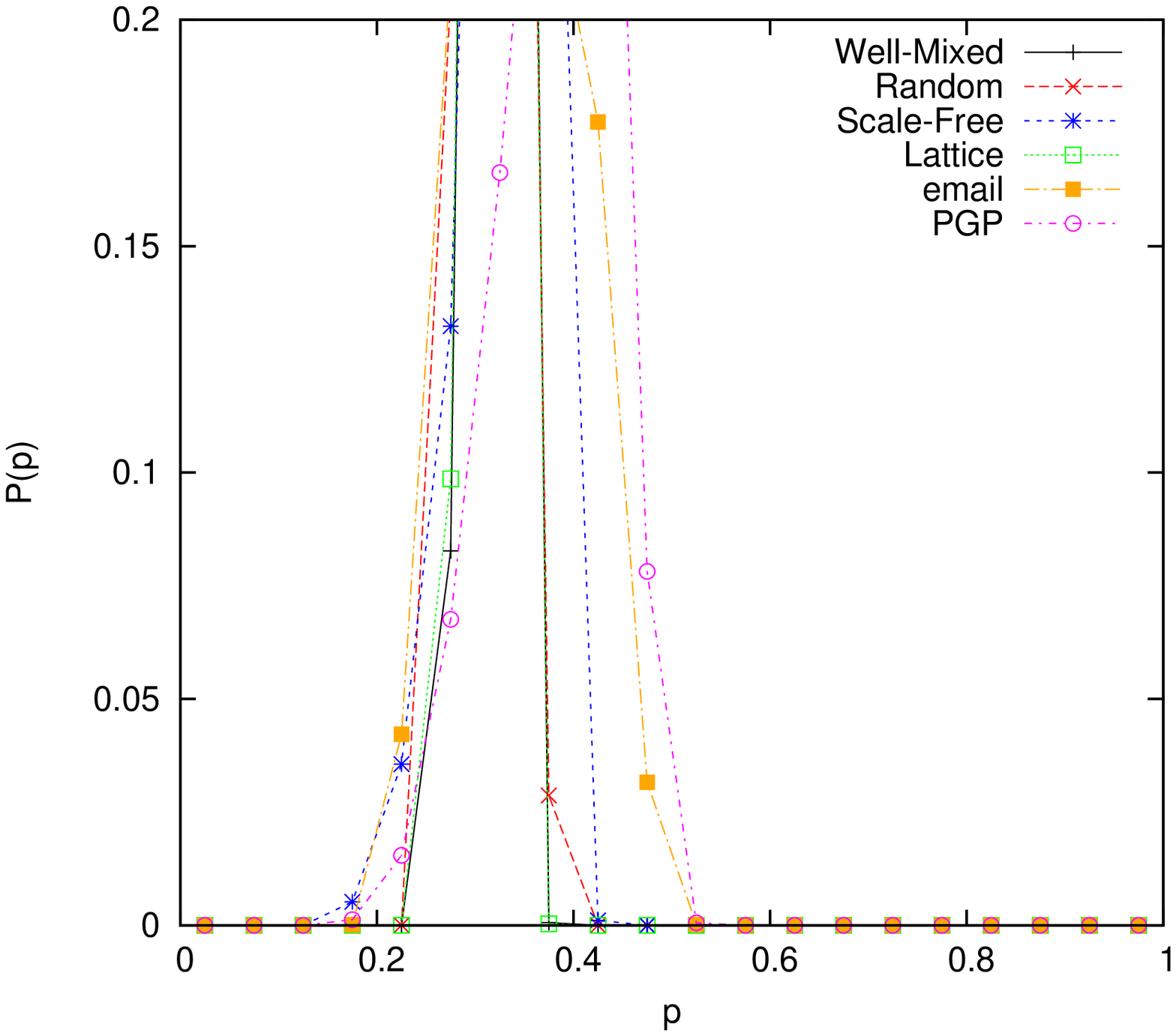}
\caption{Evolution of the level of cooperation $c$ and stationary distribution of $p$ when the evolutionary dynamics is Reinforcement Learning with $\lambda=10^{-2}$. 
From left to right: $A=1/2$, $A=5/4$, adaptive $A$ ($A^{(0)}=1/2$, $h=0.2$). Top plots refer to $S=-1/2$, bottom plots to $S=0$. 
$T=3/2$ in all cases. Results are averaged over 10 independent realizations.}
\label{fig.RL}
\end{figure*}


\begin{thebibliography}{999}
\bibitem{footB} Results for artificial networks are robust with respect to system size and link density.
\bibitem{footUI} Unconditional imitation is the only imitative dynamics whose plausibility has not been fully excluded by experiments \cite{Grujic2014}, 
but only if around a 30\% of actions are mistaken (\emph{i.e.}, an action different from that of the neighbor with the highest payoff is chosen).
\bibitem{footL} The specific value of $\lambda$ influences the convergence time of the system to its stationary state. However, if players learn too quickly ($\lambda\sim1$) 
then the parameters $p$ change too much at each iteration to reach stationary values, as typically happens for learning algorithms.
\end{thebibliography}
\end{document}